\documentclass[10pt]{article}
\usepackage{fullpage}
\usepackage{amsmath}
\usepackage{amssymb}
\usepackage{amsfonts}
\usepackage[dvips]{epsfig}
\def\##1{{\bf #1}}
\def\=#1{\underline{\underline{#1}}}

\def\+#1{\underline{\bf #1}}
\def\*#1{\underline{\underline{\bf #1}}}

\def\r#1{(\ref{#1})}
\def\l#1{\label{#1}}
\def\c#1{\cite{#1}}

\def\le{\left(}
\def\ri{\right)}
\def\les{\left[}
\def\ris{\right]}
\def\lec{\left\{}
\def\ric{\right\}}

\def\.{\cdot}

\def\epso{\epsilon_{\scriptscriptstyle 0}}

\def\muo{\mu_{\scriptscriptstyle 0}}

\def\ko{k_{\scriptscriptstyle 0}}

\def\eps{\epsilon}

\begin{document}

\begin{center}

{\bf {\LARGE On the sensitivity of directions which support Voigt wave propagation in infiltrated biaxial dielectric materials}}

\vspace{10mm} \large

 Tom G. Mackay\footnote{E--mail: T.Mackay@ed.ac.uk}\\
{\em School of Mathematics and
   Maxwell Institute for Mathematical Sciences\\
University of Edinburgh, Edinburgh EH9 3JZ, UK}\\
and\\
 {\em NanoMM~---~Nanoengineered Metamaterials Group\\ Department of Engineering Science and Mechanics\\
Pennsylvania State University, University Park, PA 16802--6812,
USA}\\

\end{center}

\begin{abstract}

Voigt wave propagation (VWP) was considered in a porous biaxial dielectric material which was infiltrated with a material of refractive index $n_a$. The infiltrated material was regarded as a homogenized composite material in the long-wavelength regime and its constitutive parameters were estimated using the extended Bruggeman homogenization formalism.
In our numerical studies, the directions which support VWP were found to vary by as much as
 $300^\circ$  per RIU as the refractive index $n_a$ was varied.
  The sensitivities achieved were acutely dependent upon  the refractive index $n_a$ and the degrees of anisotropy and dissipation of the porous biaxial material. The
 orientations,  shapes and sizes of the particles which constitute  the  infiltrating material and the porous biaxial material exerted only a secondary influence
 on the  maximum sensitivities achieved. Also, for the parameter ranges considered, the degree of porosity of the biaxial material  had little effect on
 the  maximum sensitivities achieved.
 These numerical findings bode well for the possible harnessing of VWP for optical sensing applications.

\end{abstract}

\vspace{5mm} \noindent  {\bf Keywords:} Voigt waves, extended Bruggeman
homogenization formalism, optical sensing

\section{Introduction}

Voigt wave propagation (VWP) is an example of a singular form of optical propagation.
In the nonsingular case of optical propagation in a
linear, homogeneous, anisotropic, dielectric material,  two
independent plane waves, with orthogonal polarizations and different
phase speeds, generally propagate in a given direction \c{BW}. However, in certain
dissipative biaxial dielectric materials, for example,  there exist particular directions along
which these two waves coalesce to form a single plane wave, namely a Voigt wave
 \c{Khap,Voigt,Panch,Fedorov,Agranovich,Grech}. Most conspicuously, the amplitude of this Voigt wave
 varies linearly with propagation
direction. More complex materials, such as
 bianisotropic materials, offer greater scope for VWP \c{Lakh_helicoidal_bianisotropic_98,Berry}, but
we restrict our attention here to the relatively simple case involving biaxial
dielectric materials.

While VWP constitutes a fundamental phenomenon in the optics
of anisotropic (and bianisotropic) materials, these waves have yet to be
exploited in technological applications. However, recent advances relating to engineered composite materials may lead to
 VWP being more  readily harnessed  for
 technological applications.
 For example,
 homogenized composite materials (HCMs) may be conceptualized
 which support VWP, with these HCMs  being derived from  relatively commonplace
 component materials which do not themselves support VWP \c{ML03,ML_WRM}. Furthermore, by judicious design of the homogenization process, the directions in the HCM which support VWP may be controlled \c{M2011_JOPA}.

The present study is motivated by the prospect harnessing VWP for optical sensing applications.
Specifically,  a porous biaxial dielectric material is considered; and we  investigate how the
directions  which support VWP vary as the porous material is infiltrated by a material of refractive index
$n_a$.  The infiltrated porous material is regarded as an HCM. An extended version of the Bruggeman homogenization formalism is used to estimate the HCM's constitutive parameters \c{M_JNP}. This formalism accommodates particulate component materials where the component particles may have  different shapes, orientations and sizes, and  it is not restricted to small volume fractions of the infiltrating material.

In our notation, bold typeface denotes a vector, with
the $\hat{}$ symbol signifying a unit vector. Thus,  $\hat{\#x}$, $\hat{\#y}$ and $\hat{\#z}$ represent the unit Cartesian
vectors.
  Double underlining with normal typeface denotes a  3$\times$3 dyadic; and
   $\=I = \hat{\#x} \, \hat{\#x} + \hat{\#y} \, \hat{\#y} +
\hat{\#z} \, \hat{\#z}$ is the identity 3$\times$3 dyadic. The superscript $T$ denotes the dyadic transpose; and the dyadic operator `$\mbox{det}$' yields the  determinant. Double underlining with blackboard bold typeface
denotes a 6$\times$6 dyadic.
  The symbols $\epso$
and $\muo$ represent the
permittivity and permeability of free space, respectively, while the free-space wavenumber is given as $\ko = \omega
\sqrt{\epso \muo}$, with $\omega$ being the angular frequency.

\section{Homogenization formalism} \label{Homog_Sec}

\subsection{Component materials}

Let us consider  an HCM arising from two component
materials, labelled $a$ and $b$. Suppose that  component material $a$ is  an
 isotropic dielectric material with permittivity
dyadic $\=\eps_{\,a} = \epso \eps_a \=I$, where the relative permittivity $\eps_a = n^2_a$.
Component material $b$ is taken to be an orthorhombic dielectric material characterized by the
diagonal permittivity
dyadic \c{EAB}
\begin{equation}
\=\eps_{\,b} =
\epso \les \, \eps^x_b   \, \hat{\#x} \, \hat{\#x} +
\eps^y_b \hat{\#y} \, \hat{\#y}  + \eps^z_b \hat{\#z}
\, \hat{\#z} \, \ris.
\end{equation}
  The volume fraction of material $a$ is $f_a$ and that
 of material $b$ is $f_b$; and $f_a + f_b = 1$.

The two component materials are randomly-distributed as assemblies of generally
ellipsoidal particles. We assume that all material $a$ particles have the same
orientation and all material $b$ particles have the same
orientation; in general these two orientations are not the same. However, for
simplicity,  let both material $a$ and material $b$ particles  have
the same shape.
The vector
\begin{equation} \l{er}
\#r_\ell = \eta_\ell \, \=U_{\,\ell} \. \hat{\#r}, \qquad (\ell = a,
b)
\end{equation}
prescribes
the surfaces of the
component particles, relative to their centres.
In Eq.~\r{er}, $\hat{\#r}$ is the radial  vector which
prescribes the surface of
the unit sphere. The surface of the unit sphere is mapped onto
the surface of an ellipsoid by
 the  real-symmetric surface dyadic $\=U_{\,\ell}$, while  $\eta_\ell >
0$ provides a linear measure of the ellipsoidal particle size. In order to be consistent with the notion of homogenization, it essential that  $\eta_\ell$ is much smaller than the
wavelengths involved. However, we implement an  extended version of the Bruggeman formalism for which
 $\eta_\ell$ is not required to be vanishingly small \cite{M_JNP}. We make the  assumption that
 $\eta_a \equiv \eta_b$; accordingly, in the following we write $\eta$ instead of $\eta_\ell$ $ (\ell = a,
b)$.

We choose the orientation of our coordinate system such that
the semi-major axes of the material $b$ particles are  aligned with the coordinate axes. Relative to the material $b$ particles,
 the material
$a$   particles are rotated  in the $xy$ plane
by an angle $\varphi$ with respect  to the $x$ axis.
Consequently the surface dyadic for
material $b$  is given as
\begin{equation}
\=U_{\,b} = \frac{1}{\sqrt[3]{U_x U_y U_z }} \le U_x \hat{\#x} \,
\hat{\#x} + U_y  \hat{\#y} \, \hat{\#y} + U_z \hat{\#z} \, \hat{\#z}
\ri  , \qquad (U_x, U_y, U_z > 0) ,
\end{equation}
while
  the surface dyadic for material $a$  is given as
\begin{equation}
\=U_{\,a} = \frac{1}{\sqrt[3]{U_x U_y U_z }} \, \=R_{\,z} (\varphi) \.
\le U_x \hat{\#x} \, \hat{\#x} + U_y  \hat{\#y} \, \hat{\#y} + U_z
\hat{\#z} \, \hat{\#z} \ri   \. \=R^T_{\,z} (\varphi), \qquad
(Ux, U_y, U_z > 0),
\end{equation}
where the orthogonal rotation dyadic
\begin{equation}
\=R_{\,z} (\varphi) =
 \cos \varphi
\le  \, \hat{\#x} \, \hat{\#x} + \hat{\#y} \, \hat{\#y} \, \ri +
 \sin \varphi
\le  \, \hat{\#x} \, \hat{\#y} - \hat{\#y} \, \hat{\#x} \, \ri +
\hat{\#z} \, \hat{\#z}.
\end{equation}

\subsection{Homogenized composite material}

In general the alignments of the semi-major axes of the ellipsoidal particles comprising
component materials $a$ and  $b$  are not the same. Therefore, the resulting
HCM is a biaxial dielectric material with a symmetric permittivity
dyadic of the form \cite{MW_biax1,MW_biax2}
\begin{equation}
\=\eps_{\,HCM} = \epso \les \, \eps_x   \, \hat{\#x} \, \hat{\#x} +
\eps_y \hat{\#y} \, \hat{\#y}  +
 \eps_t \le  \hat{\#x} \, \hat{\#y} + \hat{\#y} \, \hat{\#x} \, \ri + \eps_z \hat{\#z}
\, \hat{\#z} \, \ris,
\end{equation}
wherein the off-diagonal element $\eps_t \neq 0$ provided that $\varphi \neq n \pi/2$ for $n \in \mathbb{Z}$.
An extended version of the Bruggeman homogenization
formalism  \cite{M_JNP} is used  to estimate $\=\eps_{\,HCM}$. A description of this formalism is provided in Appendix~1.

\section{Voigt wave propagation}

Next we turn to the prospect of  VWP in the
HCM, for all possible directions relative to the symmetry axes of the
HCM. It is convenient do so indirectly, by considering
  VWP along the $z$ axis for all
possible orientations of the HCM. Accordingly,  the  HCM
permittivity dyadic in the rotated coordinate frame \c{ML03}
\begin{eqnarray}
\={\tilde{\eps}}_{\,HCM} (\alpha, \beta, \gamma ) &=&
 \=R_{\,z}(\gamma)\.\=R_{\,y}(\beta)\.\=R_{\,z}(\alpha)\.\=\eps_{\,HCM}\.
 \=R^T_{\,z}(\alpha)\.\=R^T_{\,y}(\beta)\.\=R^T_{\,z}(\gamma) \l{HCMrot2} \\
 &\equiv& \epso \Big[\eps_{11} \,
\hat{\#x} \, \hat{\#x} + \eps_{22} \, \hat{\#y} \,  \hat{\#y} +
\eps_{33} \, \hat{\#z} \,  \hat{\#z} +  \eps_{12} \, \le \hat{\#x}
\, \hat{\#y} + \hat{\#y} \,  \hat{\#x} \ri \nonumber \\ &&  +
\eps_{13} \, \le \hat{\#x} \, \hat{\#z} + \hat{\#z} \,  \hat{\#x}
\ri + \eps_{23} \, \le \hat{\#y} \, \hat{\#z} + \hat{\#z} \,
\hat{\#y} \ri \Big]
 \l{HCMrot1}
\end{eqnarray}
is introduced, with
the orthogonal rotation dyadic
\begin{equation}
\=R_{\,y} (\beta)= \cos \beta \le  \, \hat{\#x} \, \hat{\#x} +
\hat{\#z} \, \hat{\#z} \, \ri +
 \sin \beta
\le  \, \hat{\#z} \, \hat{\#x} - \hat{\#x} \, \hat{\#z} \, \ri +
\hat{\#y} \, \hat{\#y},
\end{equation}
and  $\alpha$, $\beta$ and $\gamma$ being the three Euler angles
\c{Arfken}.

Voigt waves propagate along the $z$ axis of the
biaxial dielectric material described by the permittivity dyadic
\r{HCMrot2} provided that the following two necessary and sufficient conditions
are satisfied \c{GL01}:
\begin{itemize}  \item[(i)] $Y (\alpha, \beta, \gamma ) = 0\:$  and
\item[(ii)] $W (\alpha, \beta, \gamma ) \neq 0$,
\end{itemize}
where the scalars
\begin{eqnarray}
Y(\alpha, \beta, \gamma ) &=& \eps^4_{13} + \eps^4_{23} -2
\eps_{23}\eps_{33}
 \les \, 2 \eps_{12}
\eps_{13} - \le \, \eps_{11} - \eps_{22}\, \ri \eps_{23}\,\ris +
\les \le \, \eps_{11}-\eps_{22}\,\ri^2 + 4 \eps^2_{12}\,\ris \,
\eps^2_{33} \nonumber \\ && + 2 \eps_{13} \lec \, \eps^2_{23}
\eps_{13} - \les \, 2 \eps_{12}\eps_{23} + \le \, \eps_{11} -
\eps_{22} \, \ri \, \eps_{13}\,\ris \eps_{33}\,\ric
\end{eqnarray}
and
\begin{equation}
W(\alpha, \beta, \gamma ) = \eps_{12} \eps_{33} - \eps_{13}
\eps_{23}\,.
\end{equation}
Significantly, these conditions  cannot be satisfied by
isotropic or  uniaxial  dielectric materials.

\section{Numerical studies}

\subsection{Preliminaries}

Let us now investigate numerically the sensitivity of the directions which support VWP  to changes in the refractive index $n_a$ of the material which
infiltrates the porous host material $b$.
These directions are yielded by the
Euler angles
$\alpha$, $\beta$ and $\gamma$ for which $Y (\alpha, \beta, \gamma ) = 0$ and $W (\alpha, \beta, \gamma ) \neq 0$.
Before  embarking on our numerical studies, let us observe that
there is no need to consider the angular coordinate $\gamma$ in our studies
 because propagation parallel to the $z$ axis (in the
rotated coordinate system) is independent of rotation about that
axis.

In the following we focus on the case where the component materials $a$ and   $b$ are specified by the constitutive parameters
\begin{equation} \l{components}
\left.
\begin{array}{l}
n_a \in \le 1,2 \ri \\
\displaystyle{\eps_b^x = 1.5 + \delta i, \quad \eps_b^y = \tau \le 1.5 + \delta i \ri, \quad \eps_b^z = \frac{1.5 +  2\delta i}{5 \tau}}
\end{array} \right\},
\end{equation}
with $\delta$ and $\tau$ being  dissipation and  anisotropy parameters, respectively.
We consider
spheroidal component particles specified by $U_x = 1+ \rho$, $U_y= U_z = 1- \frac{\rho}{18}$, in terms of a particle eccentricity parameter $\rho$.

\subsection{HCM constitutive parameters}

The relationship between the  extended (and unextended) Bruggeman estimates of the HCM's relative permittivity
parameters $\eps_{x,y,z,t}$ and the parameters specifying the component materials is a topic which has been explored in earlier works
\c{ML03,ML_WRM,M2011_JOPA}. For convenient reference, in Appendix~2 graphs of $\eps_{x,y,z,t}$ versus volume fraction $f_a$, orientation angle $\varphi$, dimensionless  size parameter $\ko \eta$, and  eccentricity parameter $\rho$, are presented for the component materials used in the present study, as specified by Eqns.~\r{components}.

\subsection{Orientations for Voigt waves}

In this subsection we explore numerically the HCM orientations which support VWP as functions of anisotropy parameter $\tau$, volume fraction $f_a$, eccentricity parameter $\rho$, particle orientation angle $\varphi$, dissipation parameter $\delta$, and dimensionless size parameter $\ko \eta$. In particular, the sensitivity of these VWP orientations with respect to small changes in the refractive index of material $a$ is considered.

In general, for a given dissipative biaxial material,  there are two directions which support VWP. The corresponding
$\alpha$ Euler angles we write as $\alpha_{1,2}$ and the  corresponding $\beta$ Euler angles we write as $\beta_{1,2}$.
The value of $W$ corresponding to the Euler angle pair $\le \alpha_1, \beta_1 \ri$ we write as $W_1$, and
the value of $W$ corresponding to the Euler angle pair $\le \alpha_2, \beta_2 \ri$ we write as $W_2$.

\subsubsection{Anisotropy parameter}

\begin{figure}[!h]
\centering \psfull
\epsfig{file=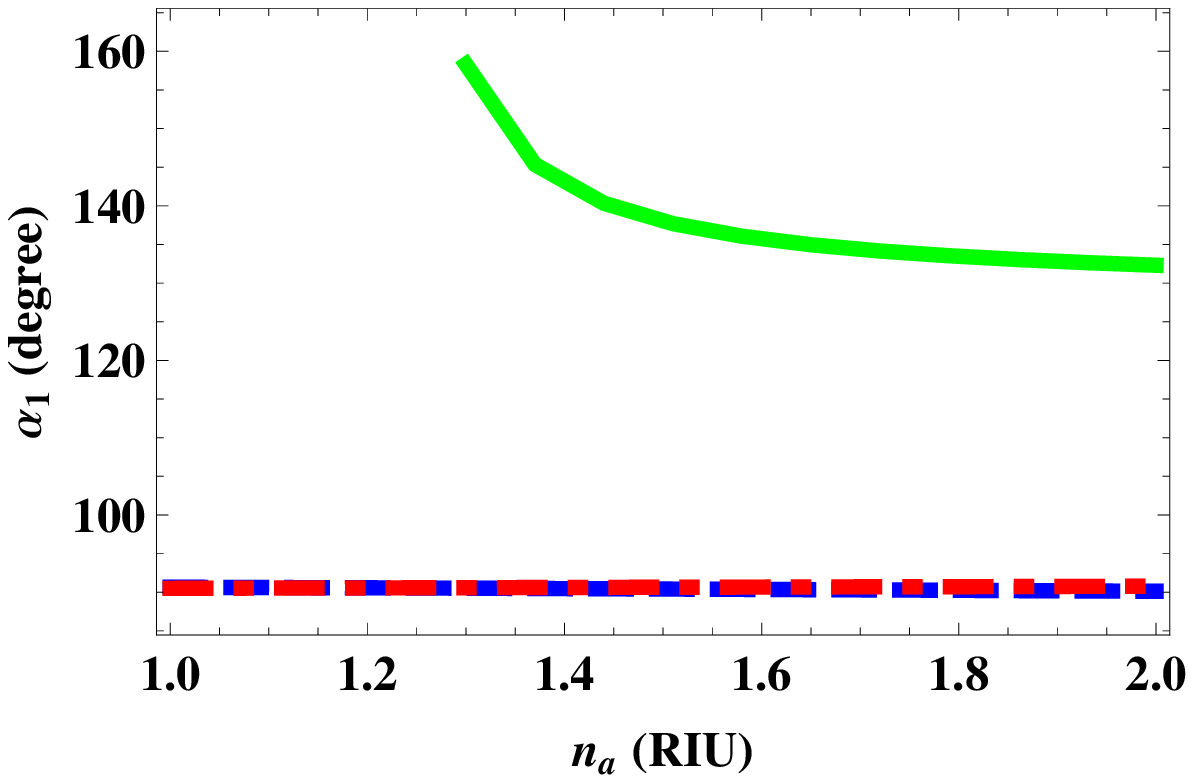,width=2.7in} \hspace{20mm}
\epsfig{file=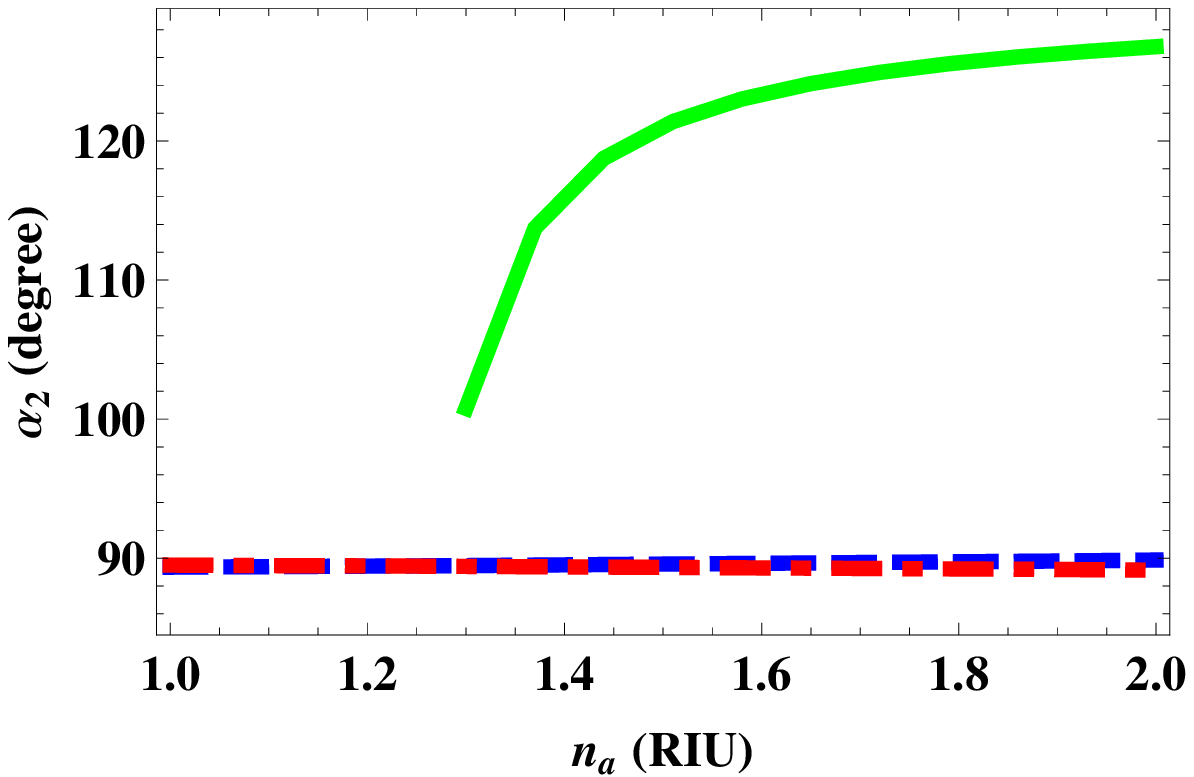,width=2.7in} \\
\epsfig{file=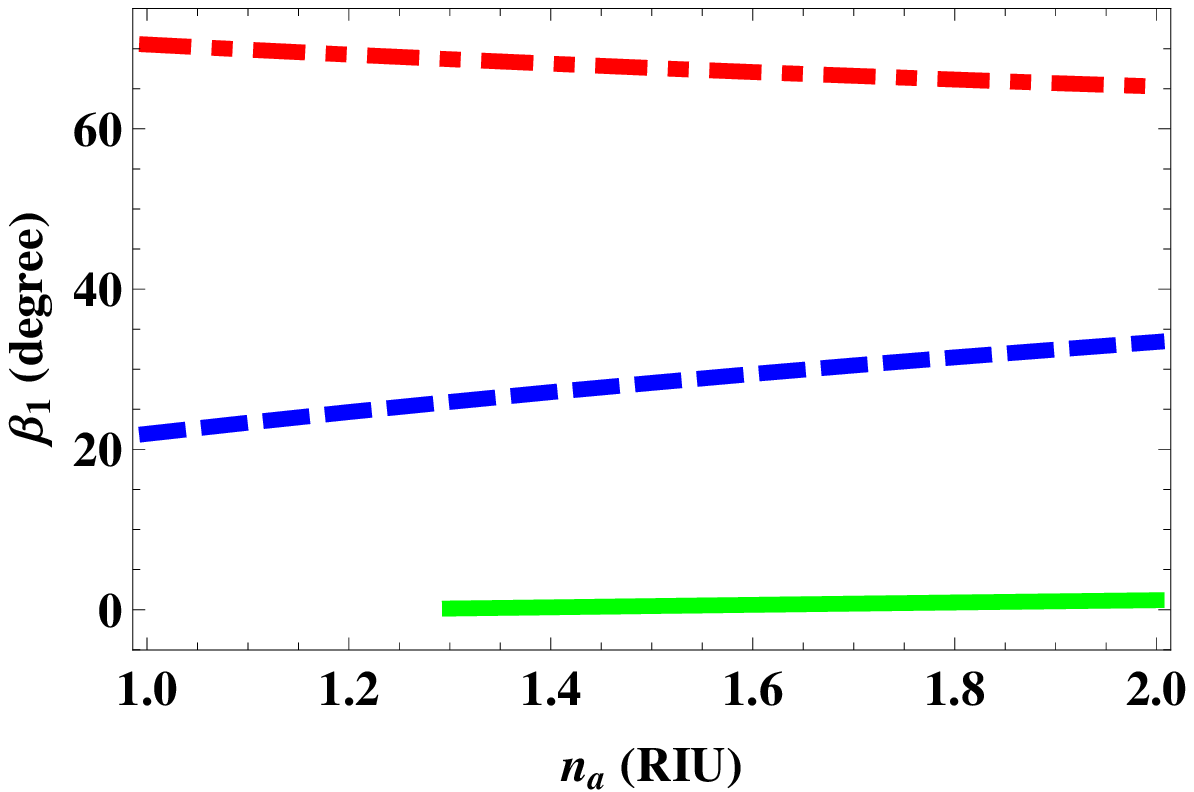,width=2.7in} \hspace{20mm}
\epsfig{file=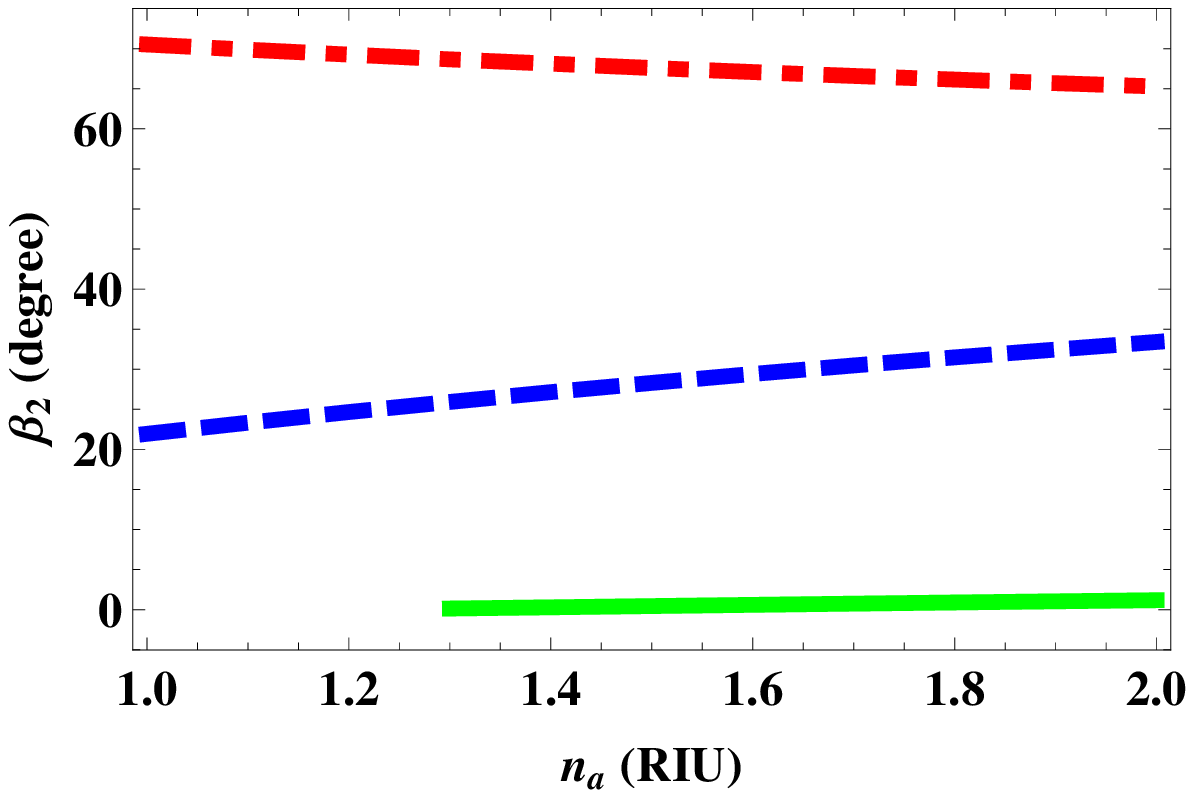,width=2.7in} \\
\epsfig{file=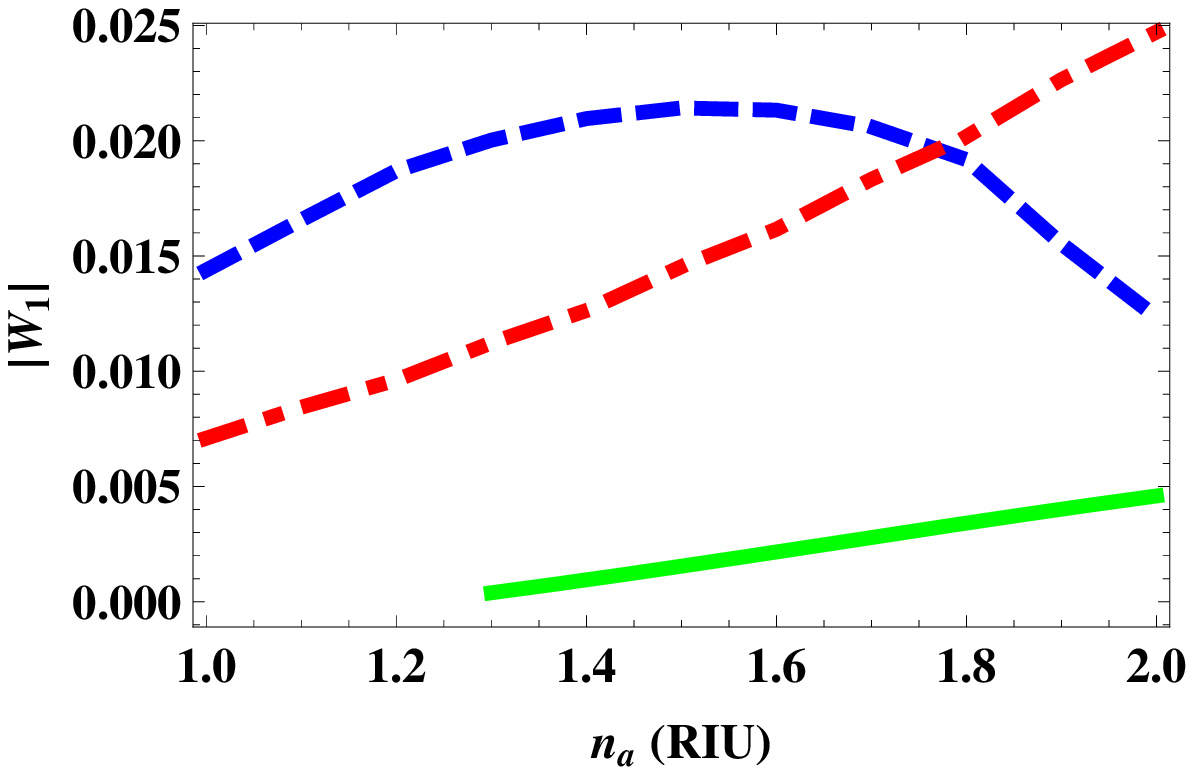,width=2.7in} \hspace{20mm}
\epsfig{file=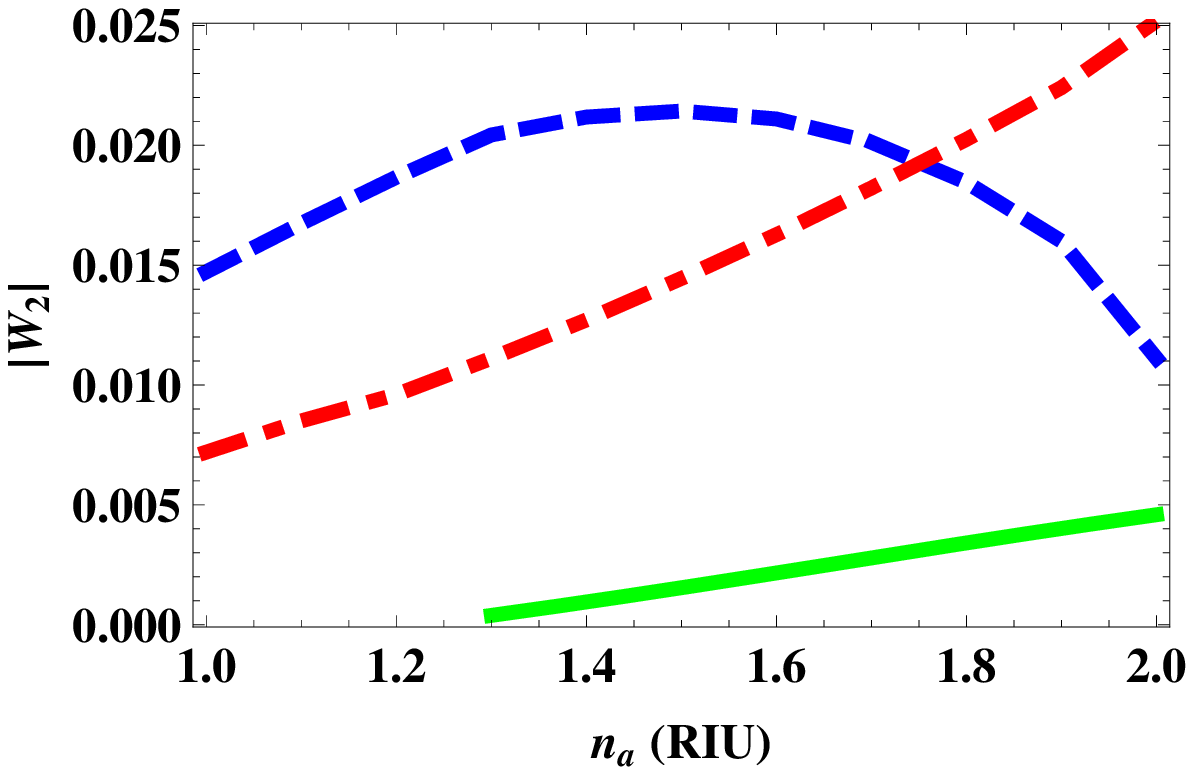,width=2.7in}
 \caption{The Euler
angles $\alpha_{1,2}$, $\beta_{1,2}$, and the  absolute value of the
quantities $W_{1,2}$, plotted versus refractive index $n_a $ for anisotropy parameter values  $\tau =  10 $ (blue, dashed curves),
$1$ (green, solid curves), and $0.1$ (red, broken dashed
curves). Values for the other component material parameters are provided in the main text. The quantities $|W_{1,2}|$ plotted
for $\tau = 1$ are scaled by 100.
 } \label{fig1}
\end{figure}

 The Euler angles
 $\alpha_{1,2}$ and $\beta_{1,2}$, together with
the corresponding values of $|W_{1,2}|$, are plotted versus $n_a \in \le
1, 2 \ri$ in Fig.~\ref{fig1} for three different values of
the anisotropy parameter:  $\tau =  10 $ (blue, dashed curves),
$1$ (green, solid curves), and $0.1$ (red, broken dashed
curves). In the case of $\tau =1$, the quantity $W$ becomes zero-valued for $n_a < 1.3$; accordingly $\tau = 1$ graphs are provided only for $n_a \in \le 1.3 , 2.0 \ri$.
For the calculations of  Fig.~\ref{fig1},
 the volume fraction $f_a = 0.25$, the
eccentricity parameter $\rho = 0.01$, the orientation angle $\varphi  = 60^\circ $, the dissipation parameter $\delta = 0.1$, and the dimensionless size parameter $\ko \eta = 0.2$.

For $\tau = 1$, the plots of $\alpha_1$ and $\alpha_2$ both vary rapidly as $n_a$ increases from 1.3, whereas the plots of $\beta_1$ and $\beta_2$ remain almost constant. The corresponding plots of $|W_{1,2}|$ vary little but most importantly these quantities are non-zero for this range of $n_a$.
For $\tau = 10$, the plots of $\beta_1$ and $\beta_2$ both increase markedly  as $n_a$ increases, whereas the plots of $\alpha_1$ and $\alpha_2$ remain almost constant. Furthermore, the plots of $\beta_1$ and $\beta_2$ are almost the same (but not exactly the same).
The corresponding values of $|W_{1,2}|$  are non-zero.
For $\tau = 0.1$ the plots of $\alpha_{1,2}$ and $\beta_{1,2}$ are qualitatively similar to those for the $\tau = 10$ case,
 except that the plots of $\beta_{1,2}$ versus $n_a$ have negative gradients.

 From the point of view of possible optical sensing applications, the sensitivity of
  the directions which support VWP
 to small changes in $n_a$  is important. Measures of these sensitivities are provided by the derivatives
  $d \alpha_{1,2} / d n_a$ and  $d \beta_{1,2} / d n_a$.
  In Fig.~\ref{fig2}
   plots of  $d \alpha_{1,2} / d n_a$ and  $d \beta_{1,2} / d n_a$
   versus refractive index $n_a $ are presented. For $\tau = 1$, the greatest sensitivities arise
   for $n_a $ values close to 1.3, where the maximum values of $| d \alpha_{1,2} / d n_a |$ are approximately $300^\circ$ per RIU.
For $\tau = 10$ and $0.1$, the greatest sensitivities arise
   for $n_a $ values close to 1.0, where the maximum values of $| d \beta_{1,2} / d n_a |$ are approximately $14^\circ$ per RIU and
$6.5^\circ$ per RIU , respectively.

In order to focus on parameter regimes which yield the greatest sensitivities, henceforth the anisotropy parameter is held constant at $\tau= 1$. Also, the quantities  $|W_{1,2}|$ have little significance apart from being non-zero; accordingly
plots of these quantities are not presented henceforth.

\subsubsection{Volume fraction}

 The Euler angles
 $\alpha_{1}$ and $\beta_{1}$, and the  derivatives   $d \alpha_{1} / d n_a$ and  $d \beta_{1} / d n_a$,  are plotted versus $n_a \in \le
1.3, 2 \ri$ in Fig.~\ref{fig3} for three different values of
the volume fraction:  $f_a =  0.1 $ (blue, dashed curves),
$0.25$ (green, solid curves), and $0.5$ (red, broken dashed
curves).
For these calculations
 the anisotropy parameter $\tau = 1$ and the other component material parameters are the same as for Figs.~\ref{fig1} and \ref{fig2}.
 The angle $\alpha_1$ varies markedly as the  volume fraction varies;
 but the degree of sensitivity, as gauged by $d \alpha_{1} / d n_a$, is essentially independent of volume fraction.
 In contrast,
 the
  angle $\beta_1$ is much less dependent upon  volume fraction;  the sensitivity measure
  $|d \beta_{1} / d n_a|$ is greater for larger volume fractions but the maximum value of this quantity is much smaller than the maximum value of the
  sensitivity measure
  $|d \alpha_{1} / d n_a|$.

 The corresponding plots for  $\alpha_2$ (not shown) are qualitatively similar to those for $\alpha_1$ except that signs of the gradients of the  graphs are reversed. The corresponding plots for  $\beta_2$ (not shown) are almost identical  to those for $\beta_1$. In fact, as these characteristics apply to all our numerical calculations, henceforth we will only present the results for $\alpha_1$ and $\beta_1$.

\subsubsection{Eccentricity parameter}

 The Euler angles
 $\alpha_{1}$ and $\beta_{1}$, and the  derivatives   $d \alpha_{1} / d n_a$ and  $d \beta_{1} / d n_a$,  are plotted versus $n_a \in \le
1.3, 2 \ri$ in Fig.~\ref{fig4} for three different values of
the eccentricity parameter:  $\rho =  0.01 $ (green, solid curves),
$1$  (blue, dashed curves), and $9$ (red, broken dashed
curves).
For these calculations
 the anisotropy parameter $\tau = 1$ and the other component material parameters are the same as for Figs.~\ref{fig1} and \ref{fig2}.
 The angle $\alpha_1$ is largely independent of the eccentricity  parameter, and likewise
  the sensitivity measure $d \alpha_{1} / d n_a$ is also largely  independent of the eccentricity parameter.
 In contrast,
 the
  angle $\beta_1$ does vary considerably as $\rho$ varies;  the sensitivity measure
  $|d \beta_{1} / d n_a|$ is generally larger for more eccentric particle shapes  however the maximum value of this quantity is much smaller than the maximum value of the
  sensitivity measure
  $|d \alpha_{1} / d n_a|$.

\subsubsection{Particle orientation angle}

The Euler angles
 $\alpha_{1}$ and $\beta_{1}$, and the  derivatives   $d \alpha_{1} / d n_a$ and  $d \beta_{1} / d n_a$,  are plotted versus $n_a \in \le
1.3, 2 \ri$ in Fig.~\ref{fig5} for three different values of
the particle orientation angle:  $\varphi =  30^\circ$ (blue, dashed curves),
$45^\circ$ (red, broken dashed
curves),  and $60^\circ$ (green, solid curves).
For these calculations
 the anisotropy parameter $\tau = 1$ and the other component material parameters are the same as for Figs.~\ref{fig1} and \ref{fig2}.
 The angle $\alpha_1$ varies considerably as $\varphi$ varies but  the sensitivity measure $d \alpha_{1} / d n_a$ is  largely  independent of $\varphi$.
 On the other hand,
 the
  angle $\beta_1$  varies very little  as $\varphi$ varies;  the sensitivity measure
  $|d \beta_{1} / d n_a|$ is slightly larger at smaller values of $\varphi$  but the maximum value of this quantity is much smaller than the maximum value of the
  sensitivity measure
  $|d \alpha_{1} / d n_a|$.

\subsubsection{Dissipation parameter}

\begin{figure}[!h]
\centering \psfull
\epsfig{file=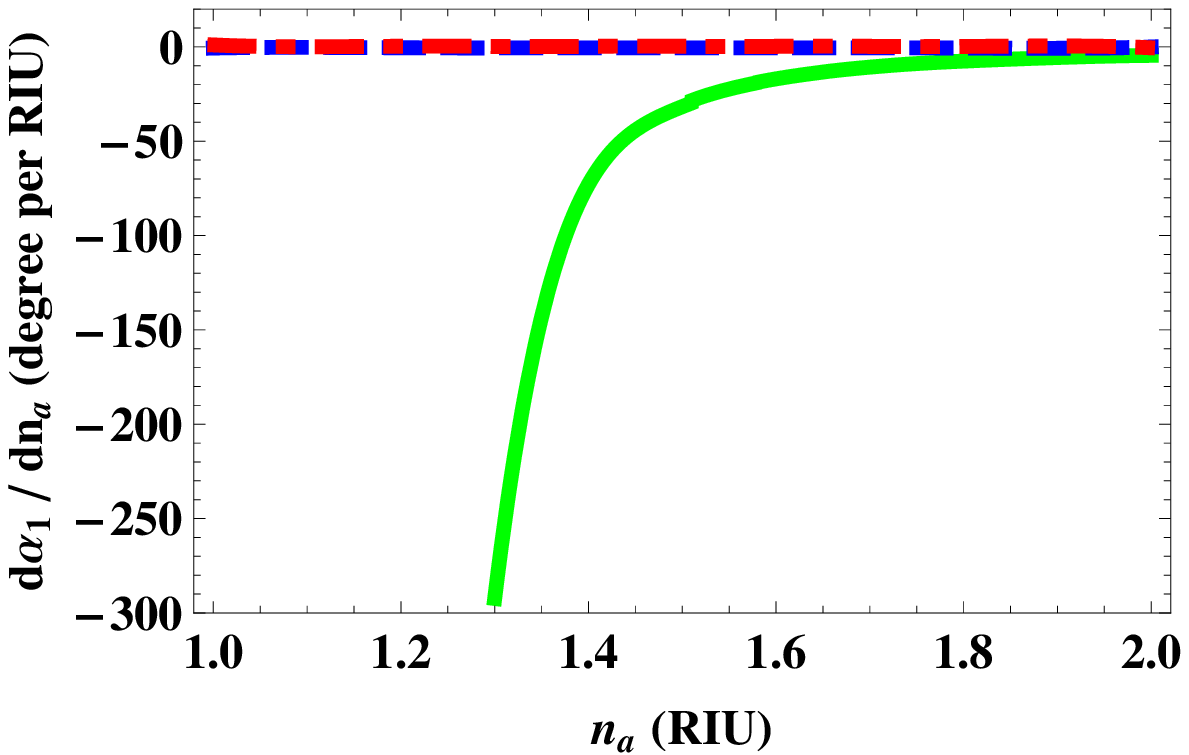,width=2.7in} \hspace{20mm}
\epsfig{file=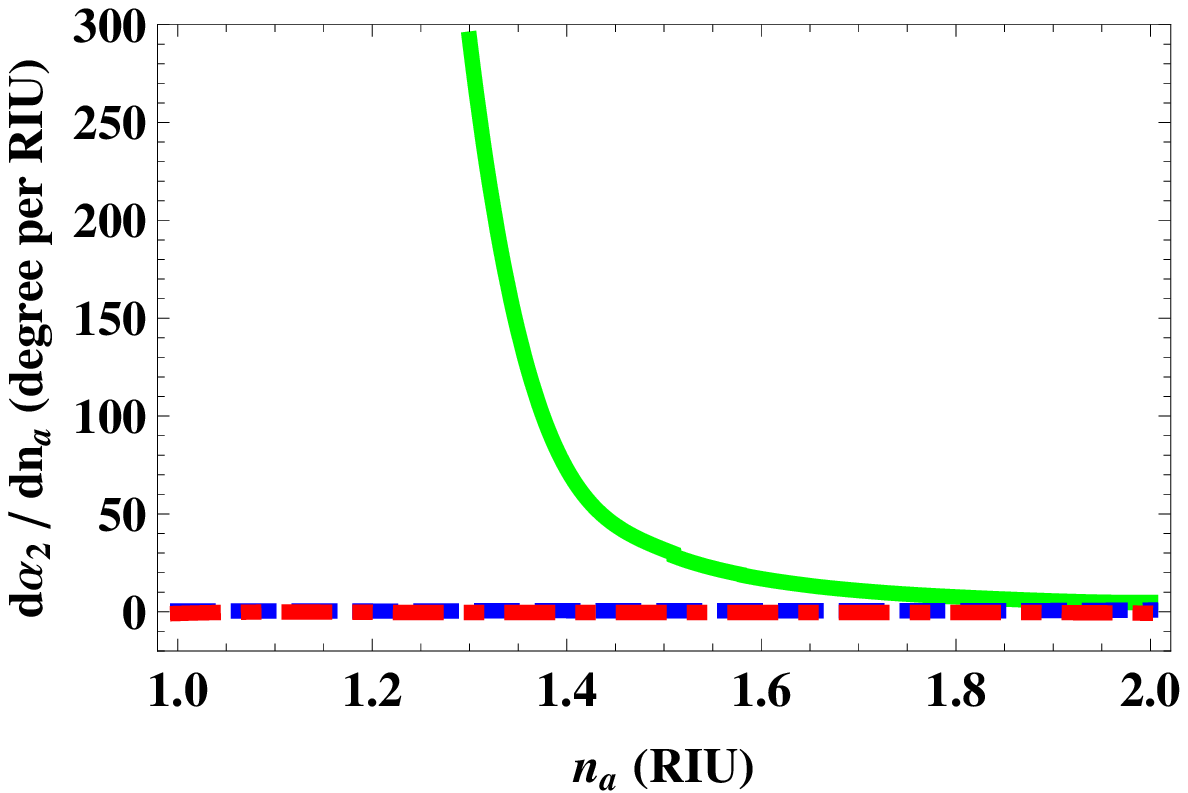,width=2.7in} \\
\epsfig{file=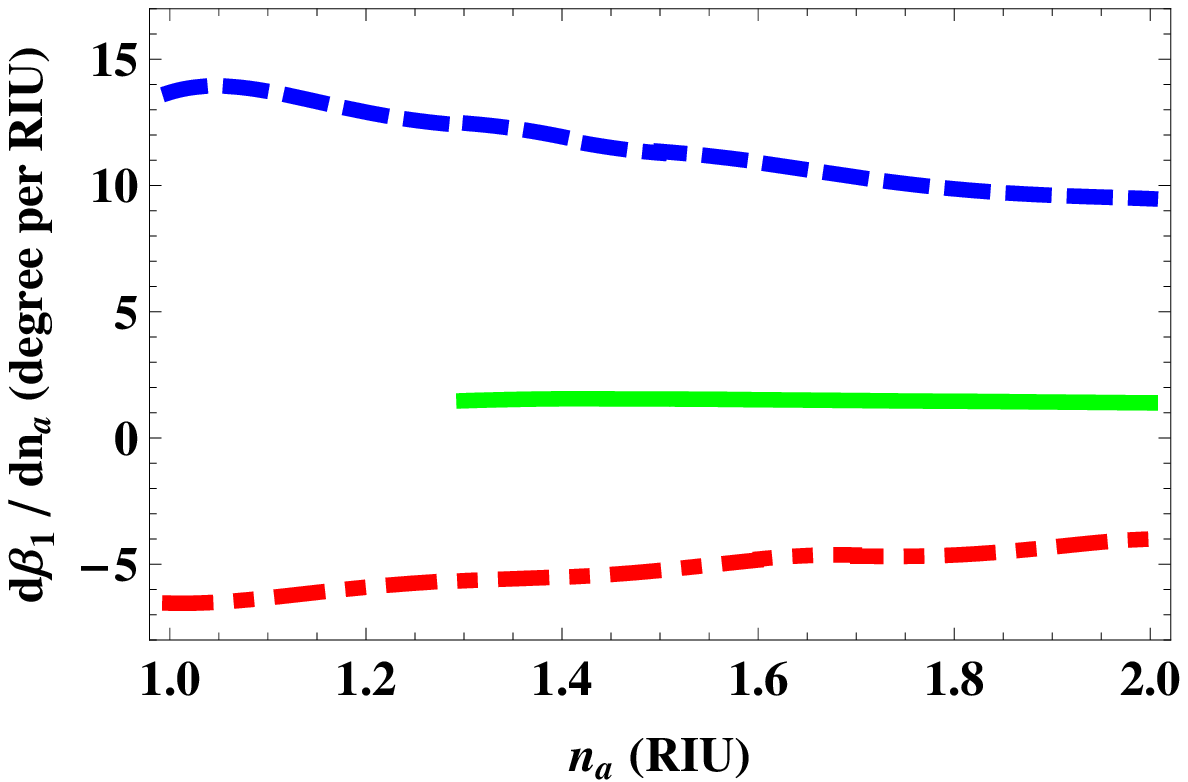,width=2.7in} \hspace{20mm}
\epsfig{file=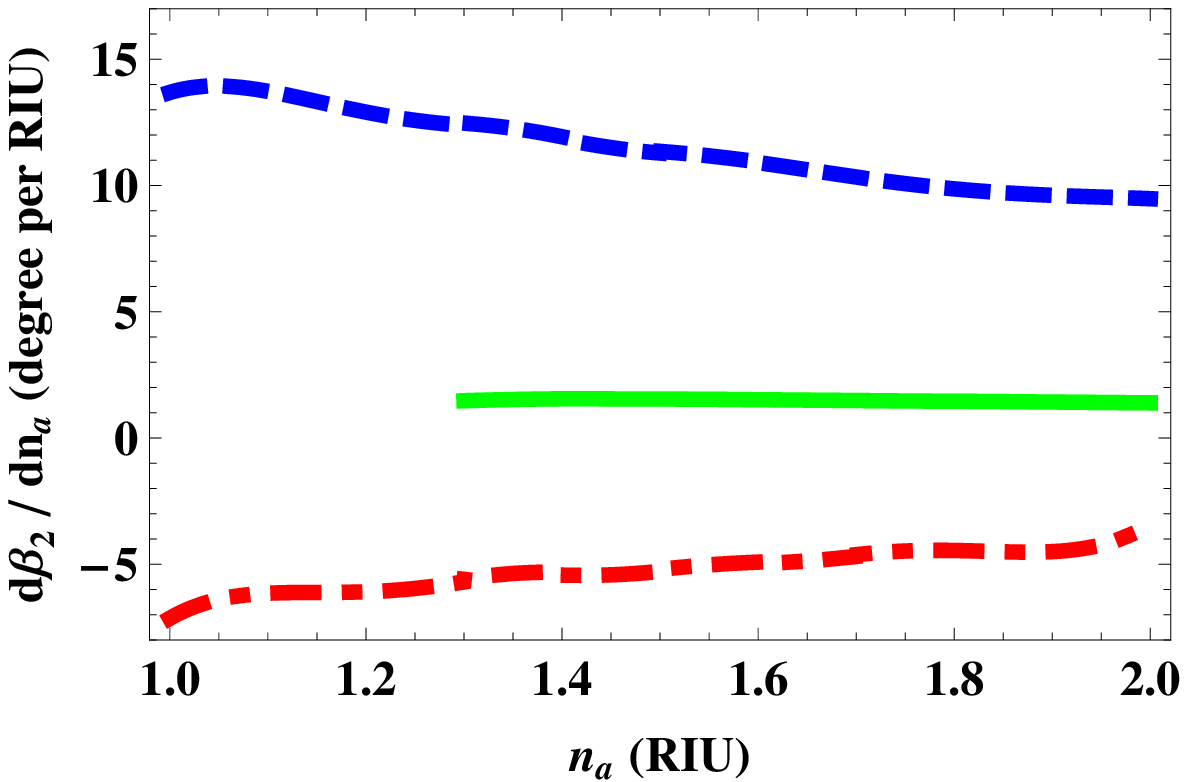,width=2.7in}
 \caption{As Fig.~\ref{fig1} but with the derivatives  $d \alpha_{1,2} / d n_a$ and  $d \beta_{1,2} / d n_a$
  plotted versus refractive index $n_a $.
 } \label{fig2}
\end{figure}

The Euler angles
 $\alpha_{1}$ and $\beta_{1}$, and the  derivatives   $d \alpha_{1} / d n_a$ and  $d \beta_{1} / d n_a$,  are plotted versus $n_a \in \le
1.3, 2 \ri$ in Fig.~\ref{fig6} for three different values of
the dissipation parameter:  $\delta =  0.01$ (blue, dashed curves),
$0.05$ (red, broken dashed
curves),  and $0.1$ (green, solid curves).
For these calculations
 the anisotropy parameter $\tau = 1$ and the other component material parameters are the same as for Figs.~\ref{fig1} and \ref{fig2}.
 The angle $\alpha_1$ varies markedly as $\delta$ varies;   the sensitivity measure $d \alpha_{1} / d n_a$ also
  varies considerably as $\delta$ varies, especially so at small values of $n_a$. We note that larger values of
  $|d \alpha_{1} / d n_a|$ are attained at larger values of $\delta$.
 On the other hand,
 the
  angle $\beta_1$  is essentially independent of the dissipation parameter;  very small changes in the sensitivity measure
  $d \beta_{1} / d n_a$ are observed as $\delta$ varies, most conspicuously at smaller values of $n_a$. As in Figs.~\ref{fig2}--\ref{fig5}, the largest values of the sensitivity measure $|d \beta_{1} / d n_a|$
  are  much smaller than the largest values of the
  sensitivity measure
  $|d \alpha_{1} / d n_a|$.

\subsubsection{Size parameter}

The Euler angles
 $\alpha_{1}$ and $\beta_{1}$, and the  derivatives   $d \alpha_{1} / d n_a$ and  $d \beta_{1} / d n_a$,  are plotted versus $n_a \in \le
1.3, 2 \ri$ in Fig.~\ref{fig7} for three different values of
the dimensionless size  parameter:  $\ko \eta =  0.2$ (green, solid curves),
$0.3$ (red, broken dashed
curves),  and $0.4$ (blue, dashed curves).
For these calculations
 the anisotropy parameter $\tau = 1$, the dissipation parameter $\delta = 0.01$,
  and the other component material parameters are the same as for Figs.~\ref{fig1} and \ref{fig2}.
 The angle $\alpha_1$ varies moderately as $\varphi$ varies, most obviously at larger values of $n_a$;
   the sensitivity measure $d \alpha_{1} / d n_a$
  varies only slightly as $\ko \eta$ varies,  most obviously at larger values of $n_a$. Slightly larger values of
  $|d \alpha_{1} / d n_a|$ are attained at smaller values of $\ko \eta$.
 In a similar vein,
 both the
  angle $\beta_1$  and $d \beta_{1} / d n_a$ vary moderately as $\ko \eta$ varies, most obviously at larger values of $n_a$.
  The largest values of $d \beta_{1} / d n_a$ are attained at larger values of $n_a$ for larger values of $\ko \eta$.
   However, as in Figs.~\ref{fig2}--\ref{fig6}, the largest values of the sensitivity measure $|d \beta_{1} / d n_a|$
  are  much smaller than the
  largest values of the sensitivity measure
  $|d \alpha_{1} / d n_a|$.

\section{Closing remarks}

For a porous biaxial dielectric  host  material infiltrated by a material of refractive index $n_a$,
our numerical studies  have revealed sensitivities of up to $300^\circ$  per RIU for the directions which support
 VWP. The sensitivities achieved are acutely dependent upon the degrees of anisotropy and dissipation of the host material,  and the refractive index $n_a$. The
 orientations,  shapes and sizes of the particles which constitute  the component materials exert only a secondary influence
 on the  maximum sensitivities achieved. Also, for the parameter ranges considered, the volume fraction has little effect on
 the  maximum sensitivities achieved.

 These numerical findings bode well for the possible harnessing of VWP for optical sensing applications. Such sensitivities of up to $300^\circ$  per RIU compare favourably to sensitivities reported for optical sensing based  on the excitation of surface-plasmon-polariton waves \c{Homola_book,AZLe,Abdulhalim}. In particular, the maximum sensitivities reported in recent studies involving surface-plasmon-polariton waves
  excited
  at the planar surfaces of sculptured thin films are an order of magnitude smaller than those found here for VWP \c{PML_JOSAB,ML_SJ_2012,SJJ_TGM_OC_2012}.

\begin{figure}[!h]
\centering \psfull
\epsfig{file=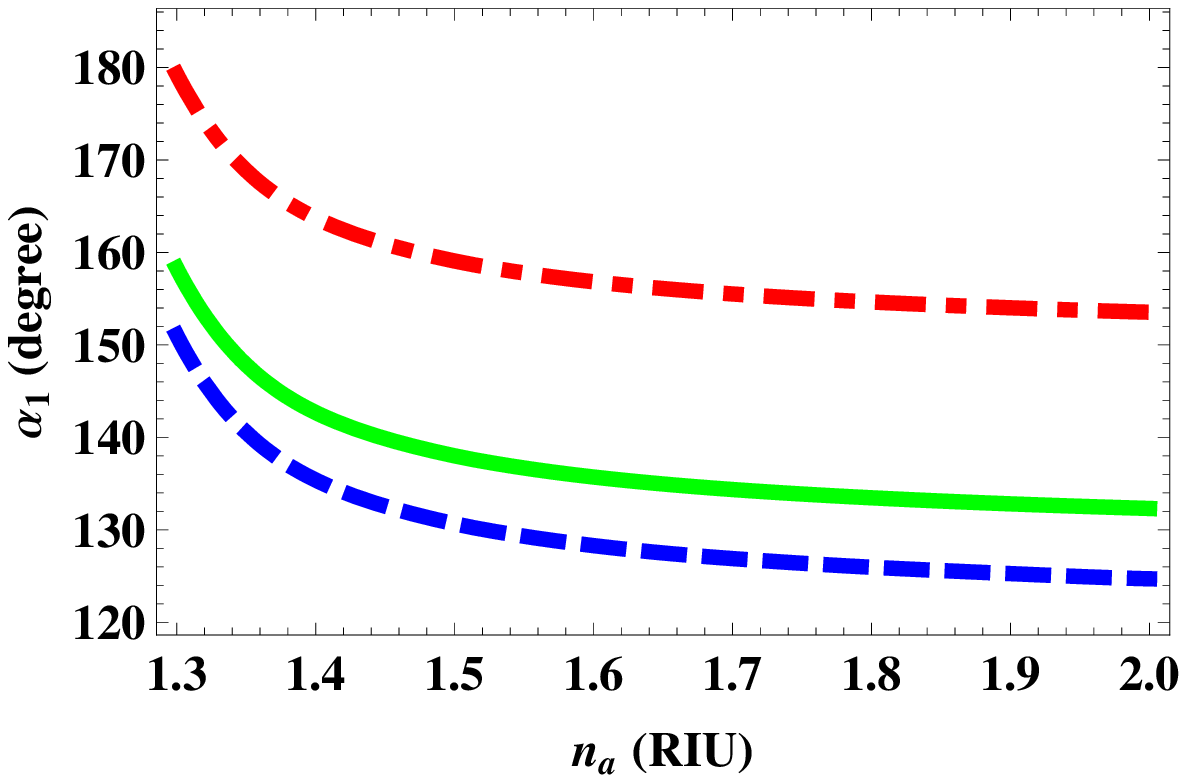,width=2.7in} \hspace{20mm}
\epsfig{file=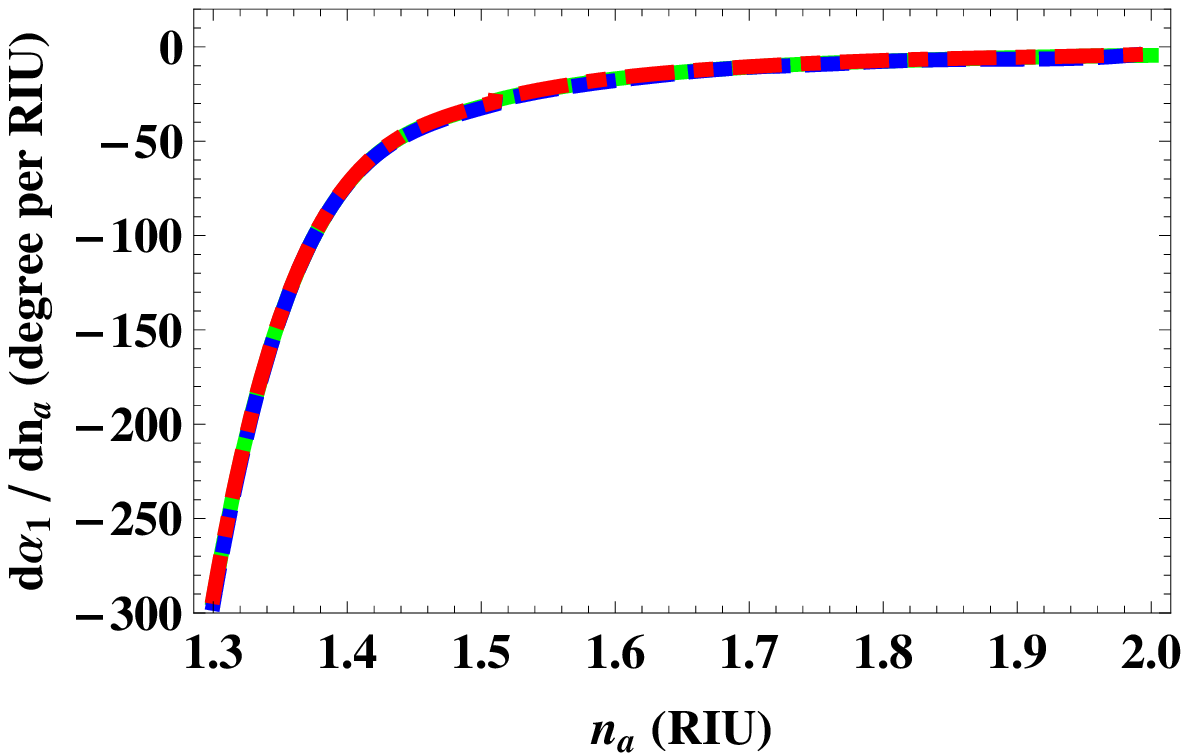,width=2.7in} \\
\epsfig{file=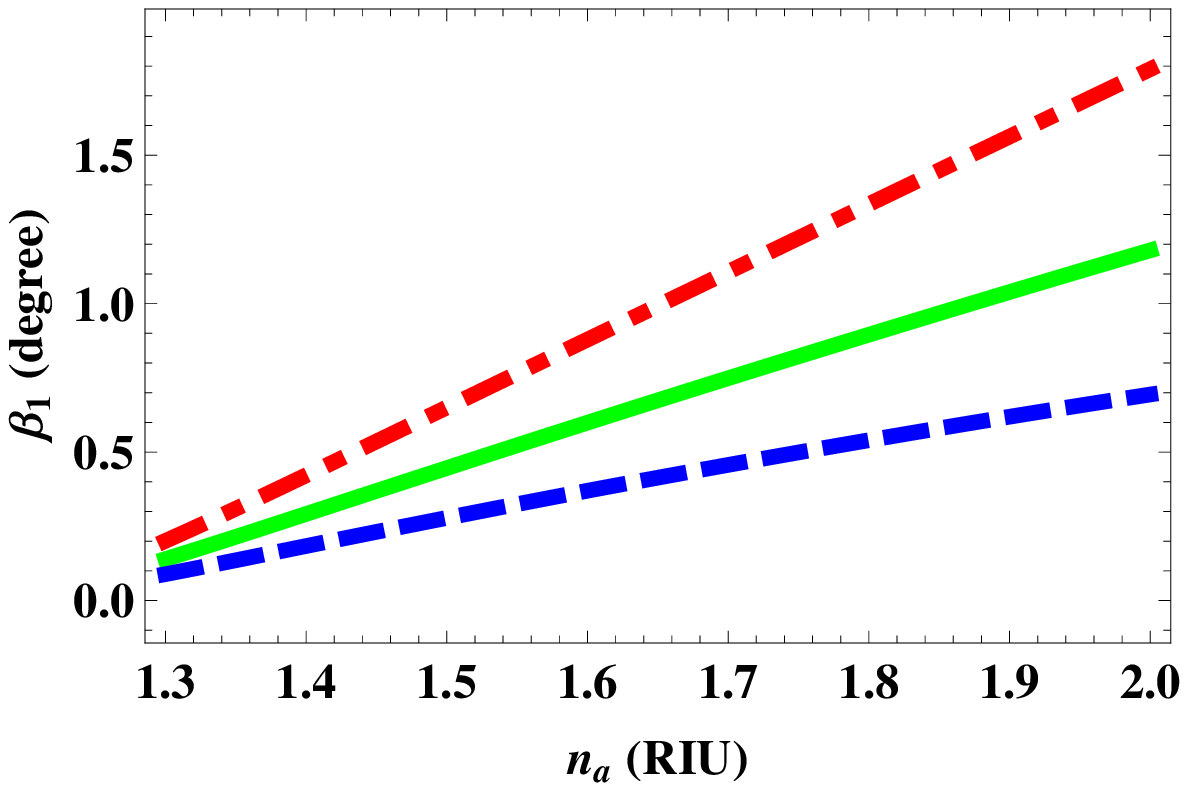,width=2.7in} \hspace{20mm}
\epsfig{file=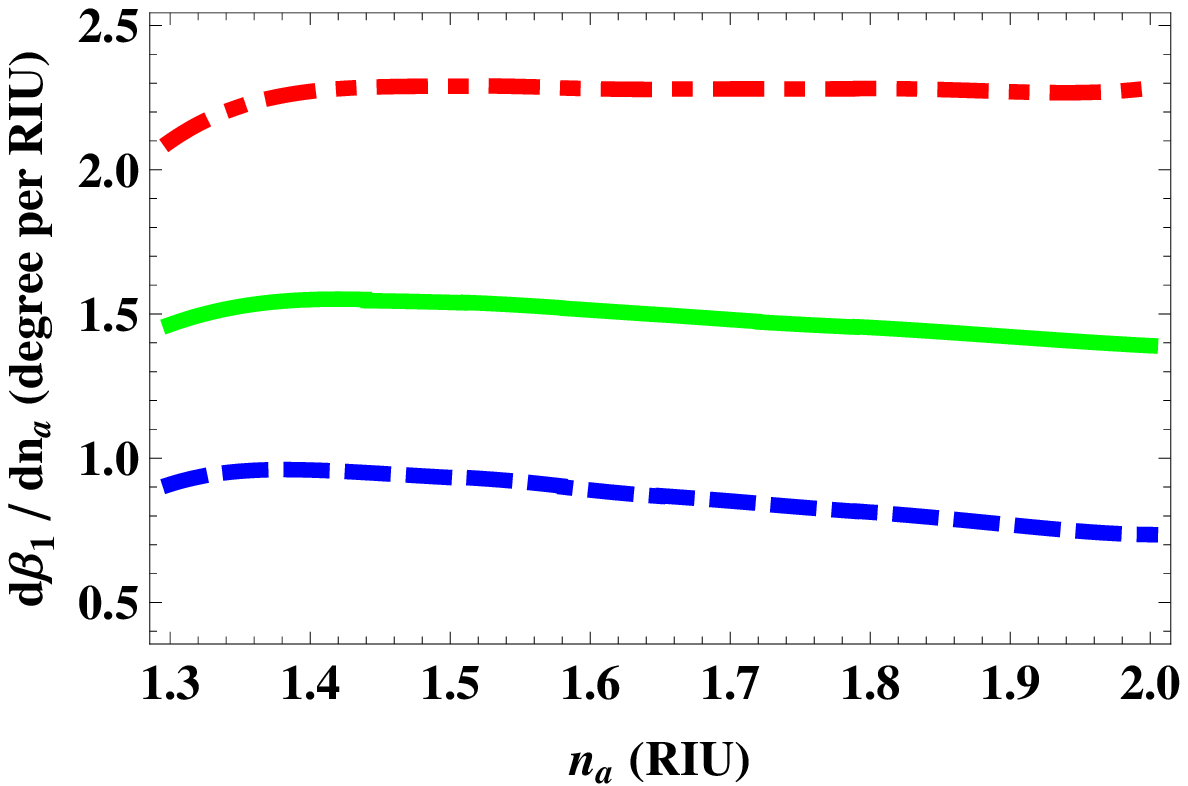,width=2.7in}
 \caption{The Euler
angles $\alpha_{1}$ and $\beta_{1}$, and the derivatives
 $d \alpha_{1} / d n_a$ and  $d \beta_{1} / d n_a$,
 plotted versus refractive index $n_a $ for volume fractions  $f_a =  0.1 $ (blue, dashed curves),
$0.25$ (green, solid curves), and $0.5$ (red, broken dashed
curves). Values for the other component material parameters are provided in the main text.
 } \label{fig3}
\end{figure}

\begin{figure}[!h]
\centering \psfull
\epsfig{file=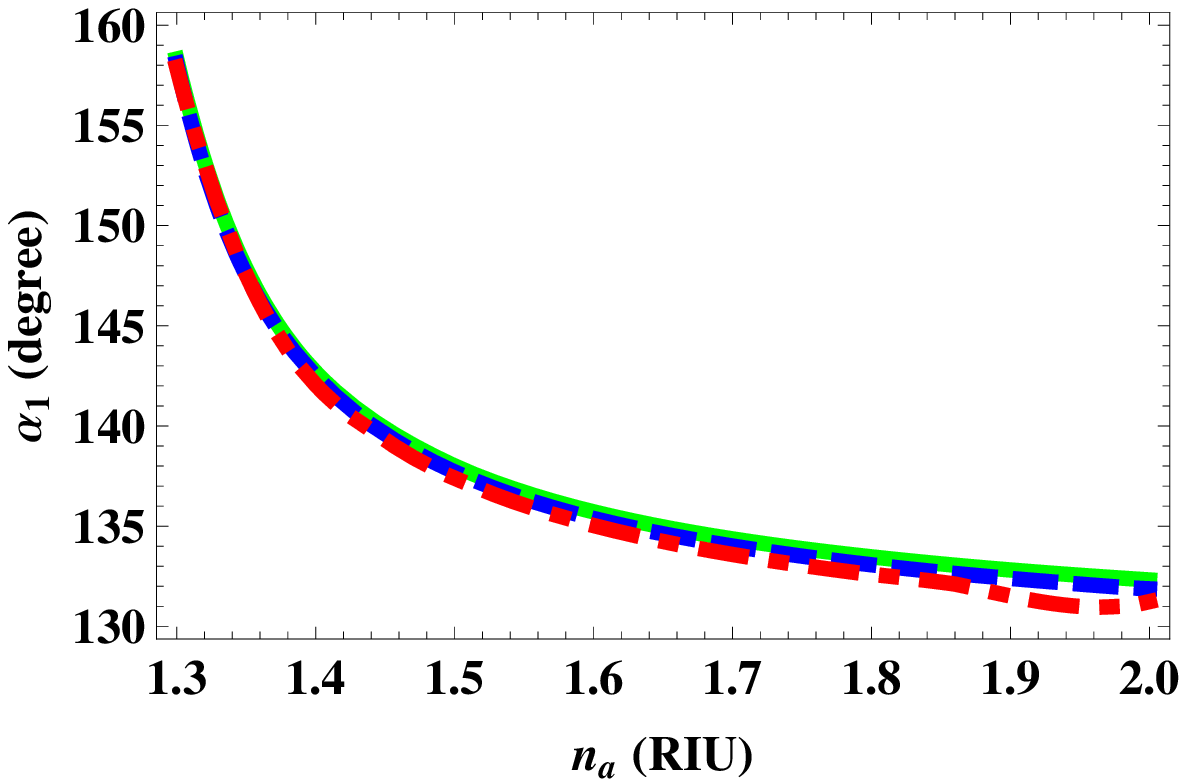,width=2.7in} \hspace{20mm}
\epsfig{file=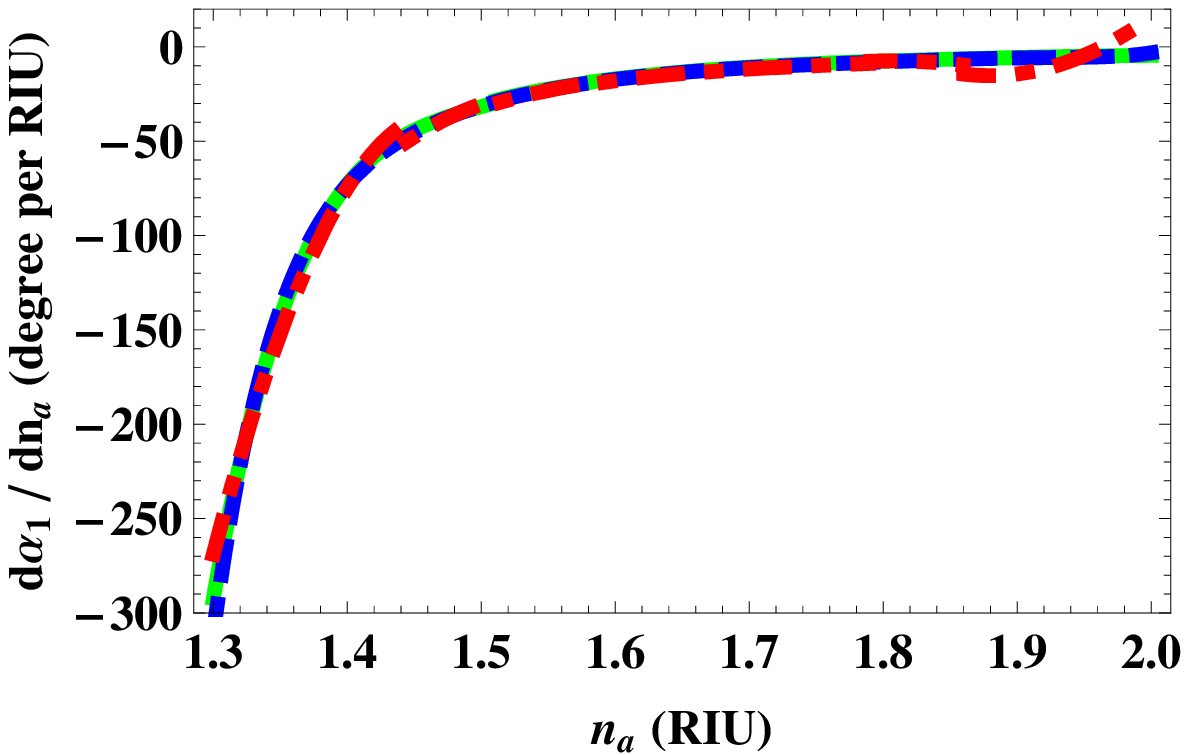,width=2.7in} \\
\epsfig{file=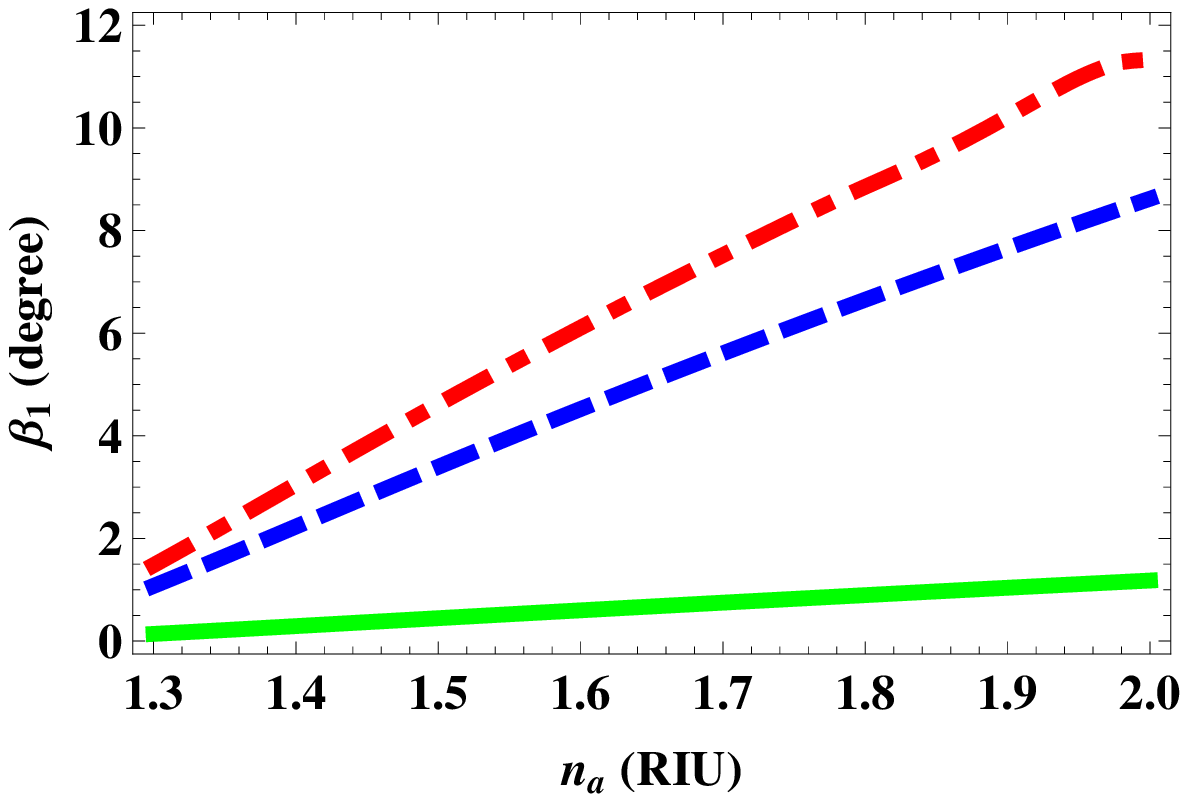,width=2.7in} \hspace{20mm}
\epsfig{file=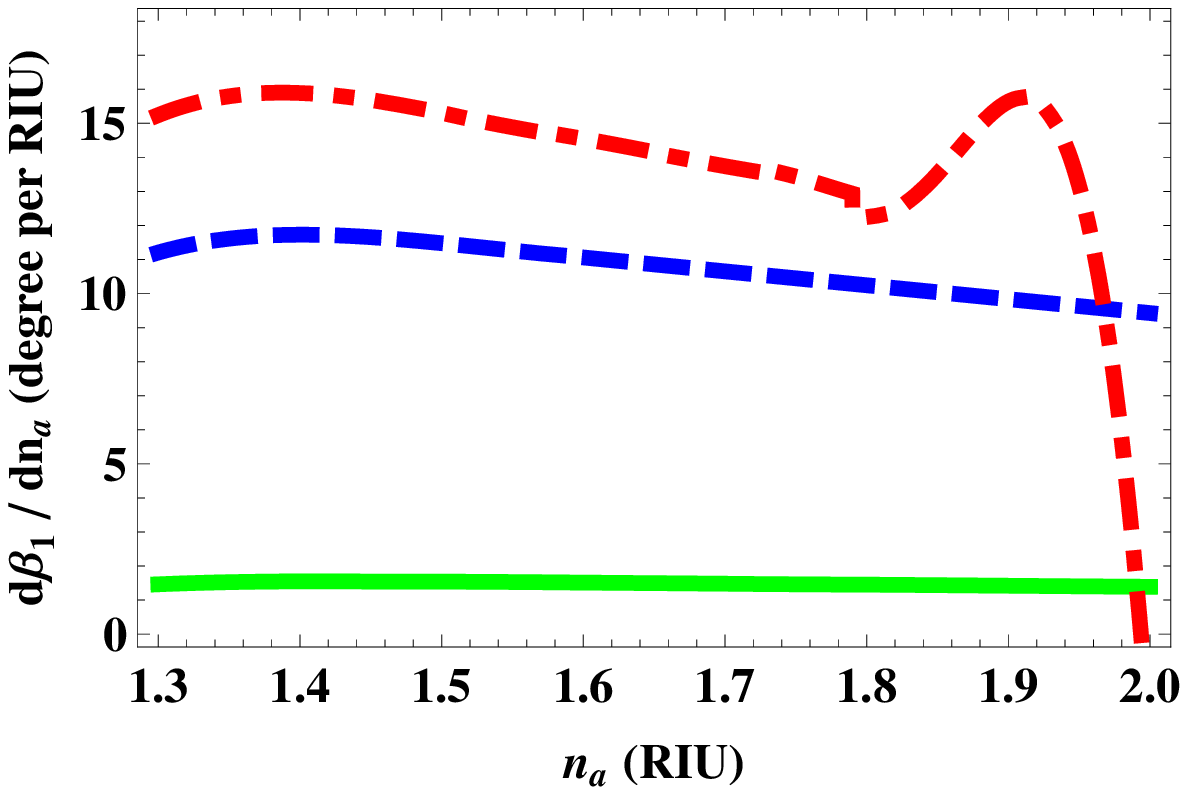,width=2.7in}
 \caption{The Euler
angles $\alpha_{1}$ and $\beta_{1}$, and the derivatives
 $d \alpha_{1} / d n_a$ and  $d \beta_{1} / d n_a$,
 plotted versus refractive index $n_a $ for eccentricity parameter values  $\rho =   0.01 $ (green, solid curves),
$1$  (blue, dashed curves), and $9$ (red, broken dashed
curves). Values for the other component material parameters are provided in the main text.
 } \label{fig4}
\end{figure}

\begin{figure}[!h]
\centering \psfull
\epsfig{file=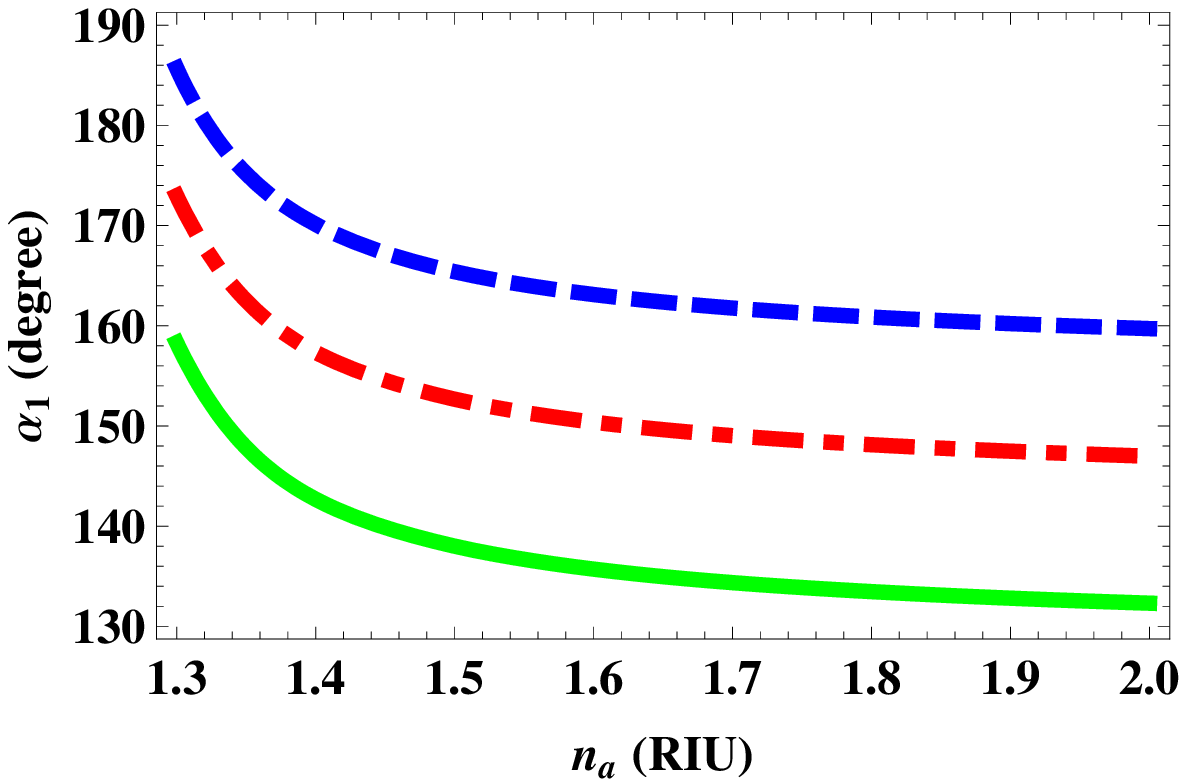,width=2.7in} \hspace{20mm}
\epsfig{file=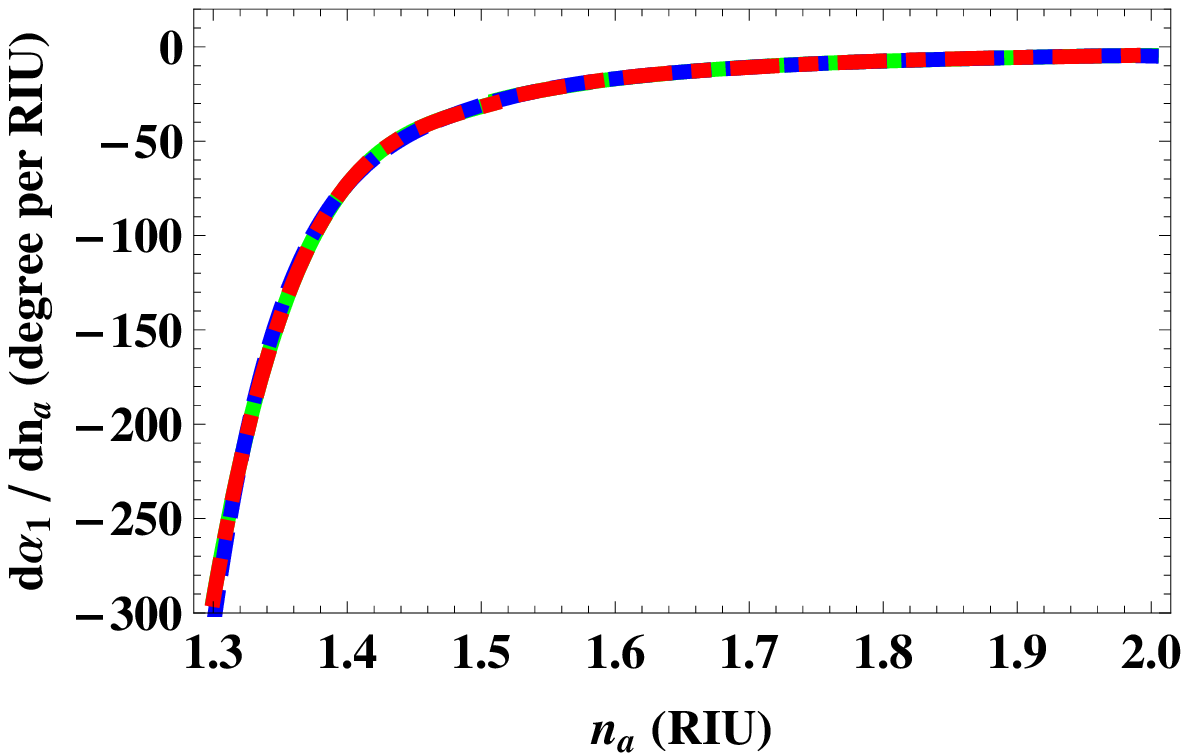,width=2.7in} \\
\epsfig{file=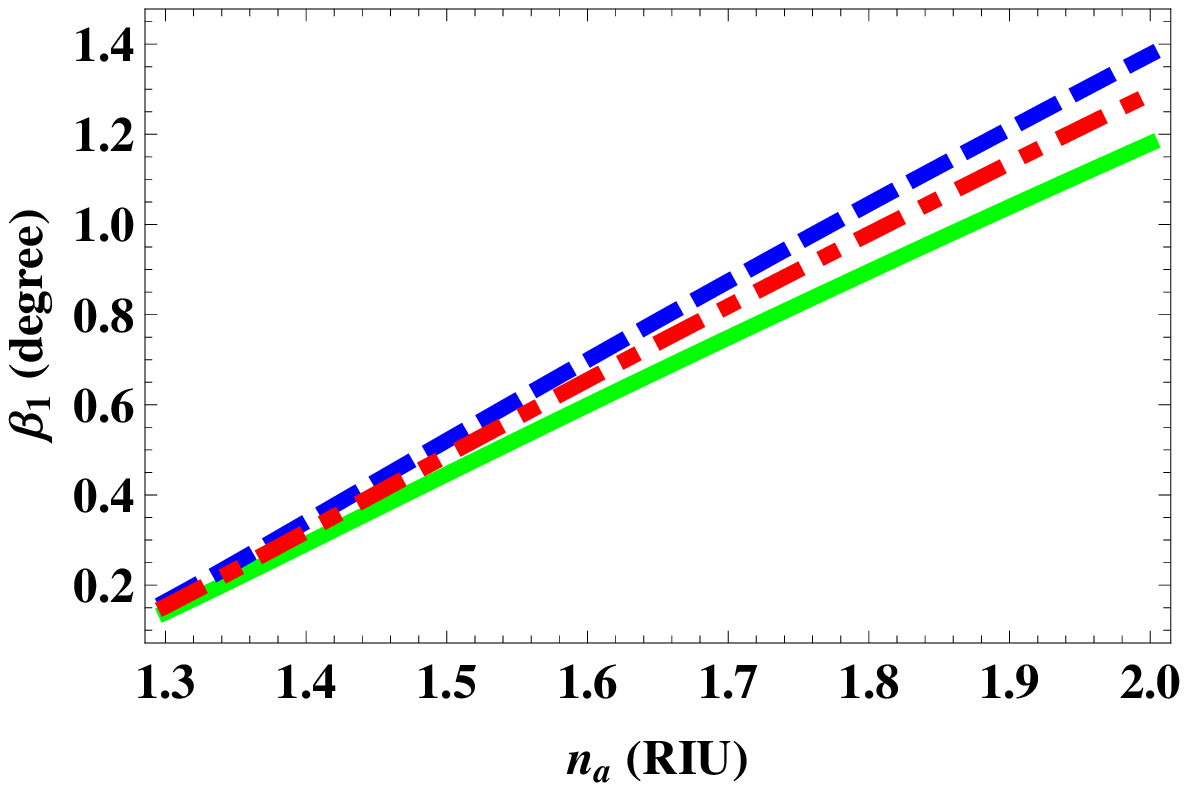,width=2.7in} \hspace{20mm}
\epsfig{file=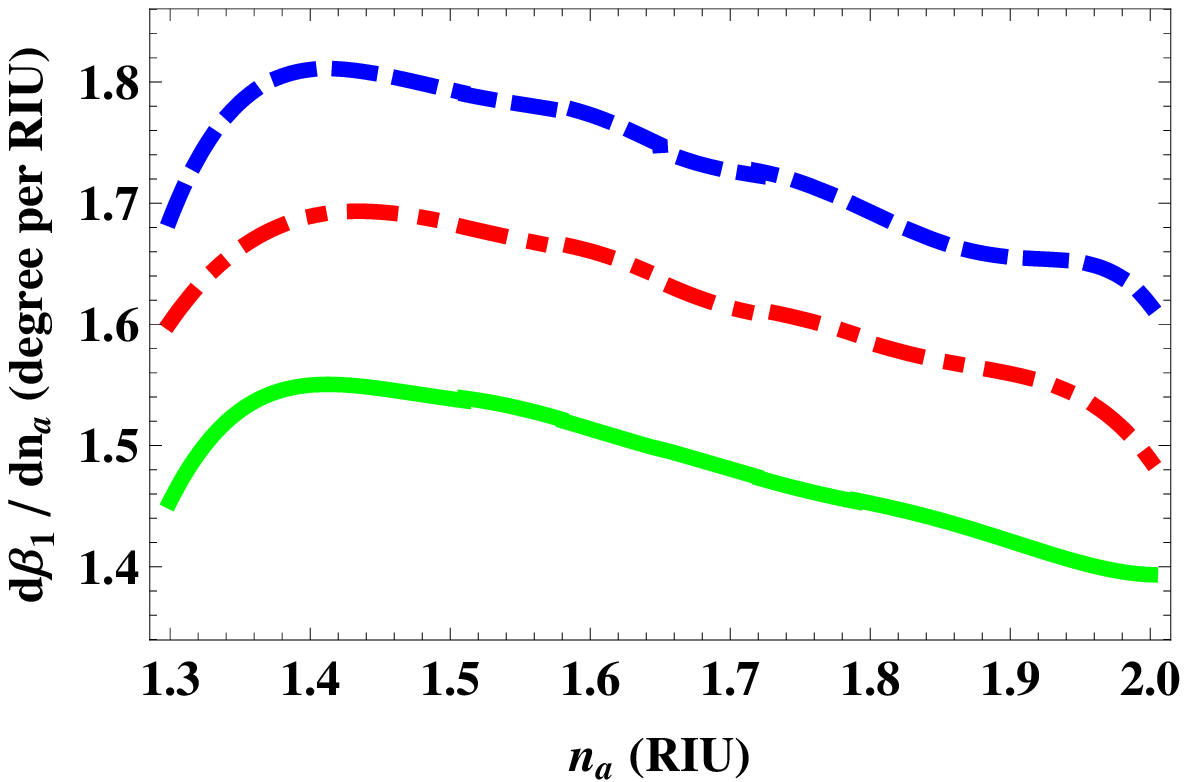,width=2.7in}
 \caption{The Euler
angles $\alpha_{1}$ and $\beta_{1}$, and the derivatives
 $d \alpha_{1} / d n_a$ and  $d \beta_{1} / d n_a$,
 plotted versus refractive index $n_a $ for particle orientation angles  $\varphi =   30^\circ $ (blue, dashed curves),
$45^\circ$   (red, broken dashed
curves), and $60^\circ$  (green, solid curves). Values for the other component material parameters are provided in the main text.
 } \label{fig5}
\end{figure}

\begin{figure}[!h]
\centering \psfull
\epsfig{file=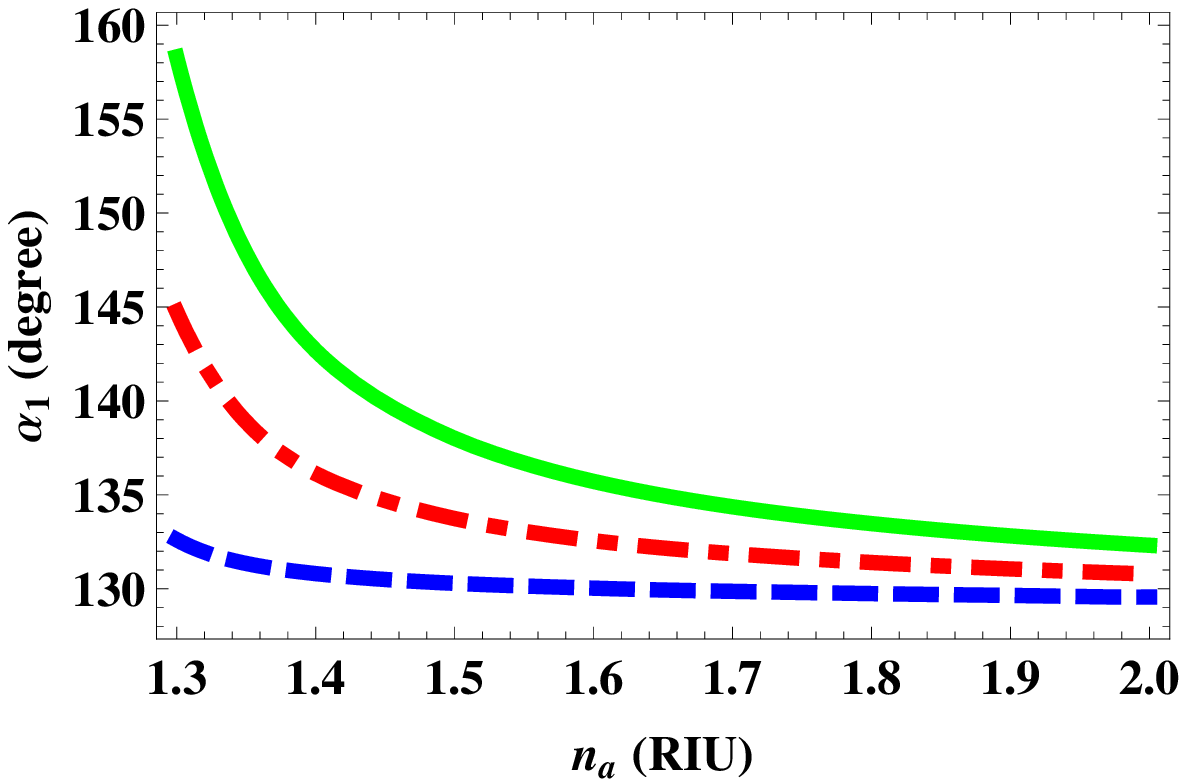,width=2.7in} \hspace{20mm}
\epsfig{file=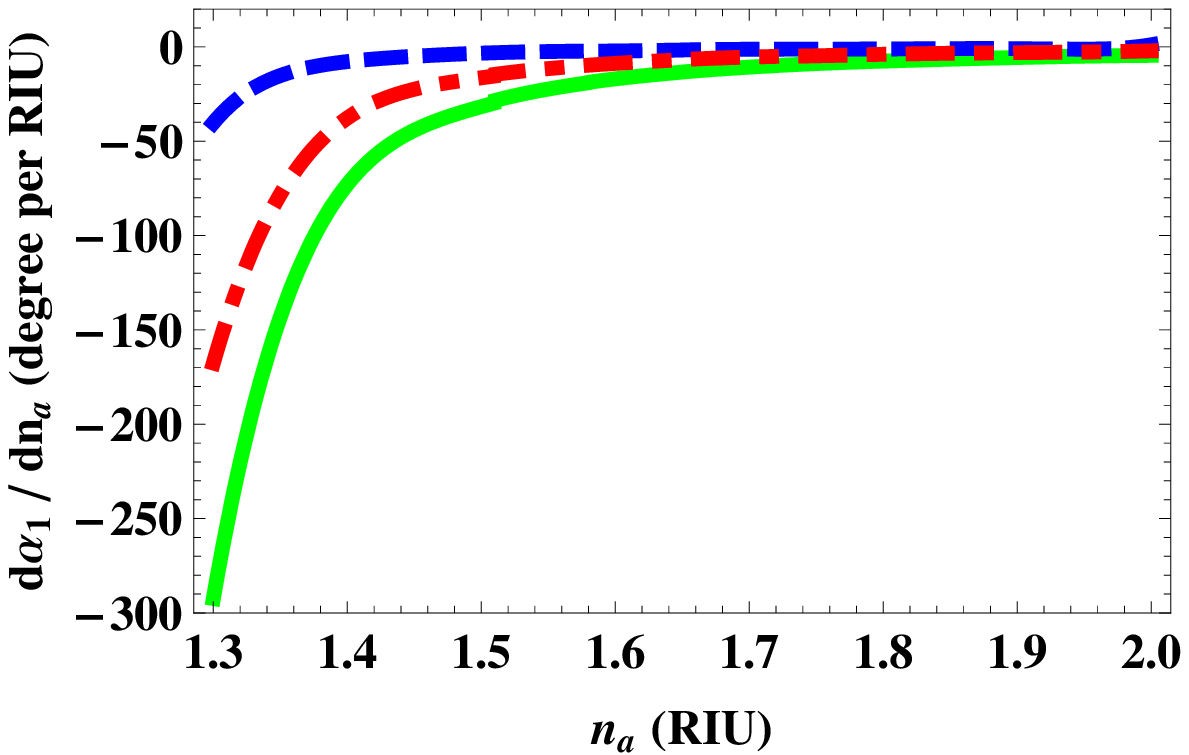,width=2.7in} \\
\epsfig{file=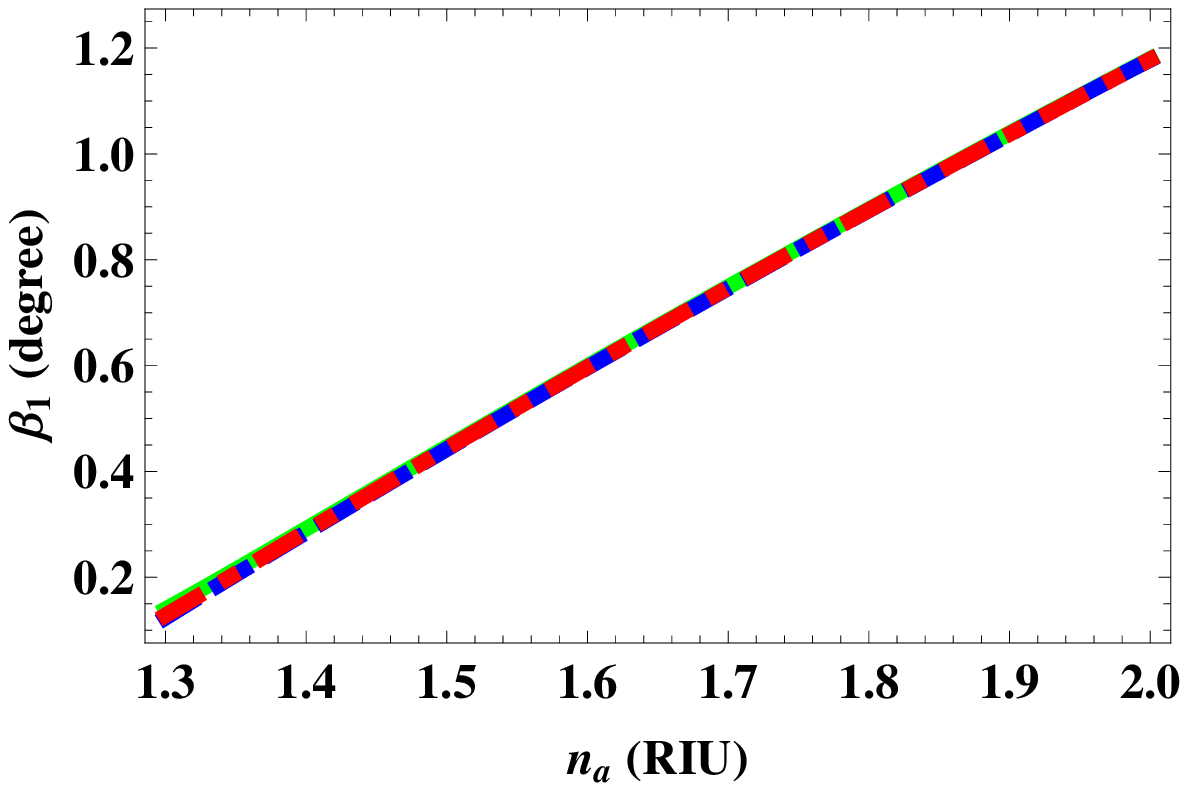,width=2.7in} \hspace{20mm}
\epsfig{file=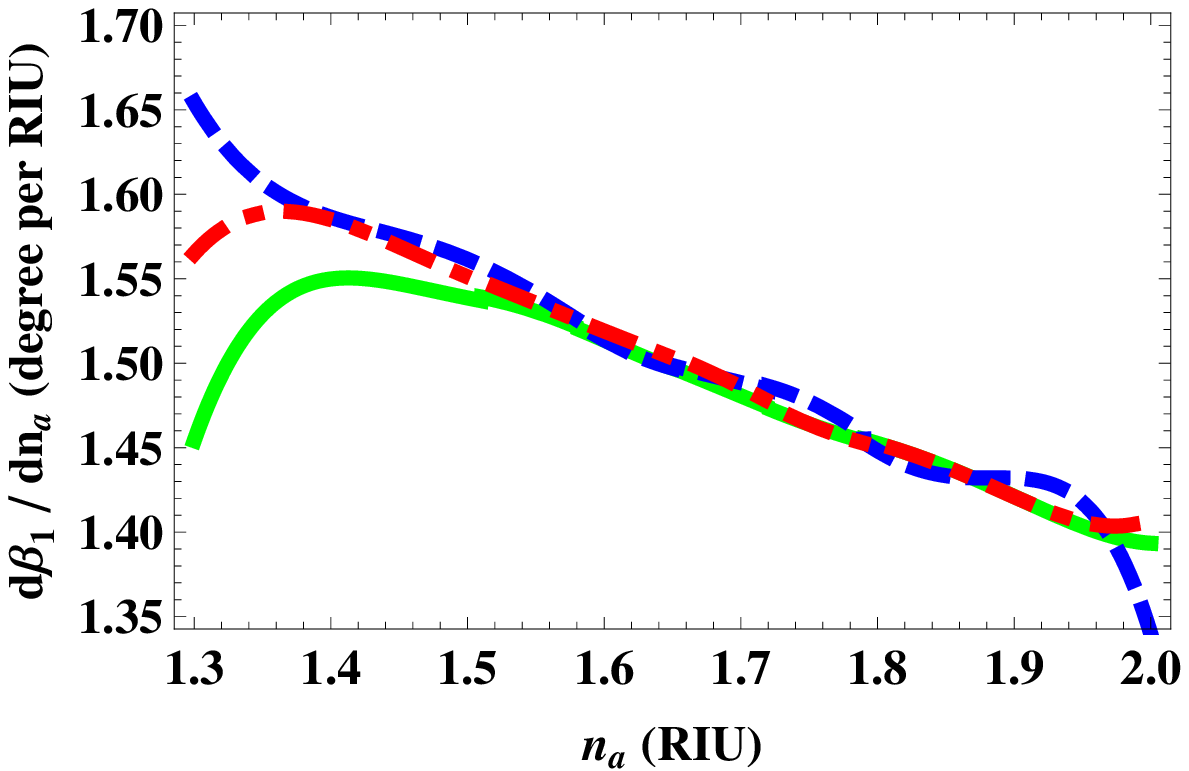,width=2.7in}
 \caption{The Euler
angles $\alpha_{1}$ and $\beta_{1}$, and the derivatives
 $d \alpha_{1} / d n_a$ and  $d \beta_{1} / d n_a$,
 plotted versus refractive index $n_a $ for dissipation parameter values  $\delta =   0.01 $ (blue, dashed curves),
$0.05$   (red, broken dashed
curves), and $0.1$  (green, solid curves). Values for the other component material parameters are provided in the main text.
 } \label{fig6}
\end{figure}

\begin{figure}[!h]
\centering \psfull
\epsfig{file=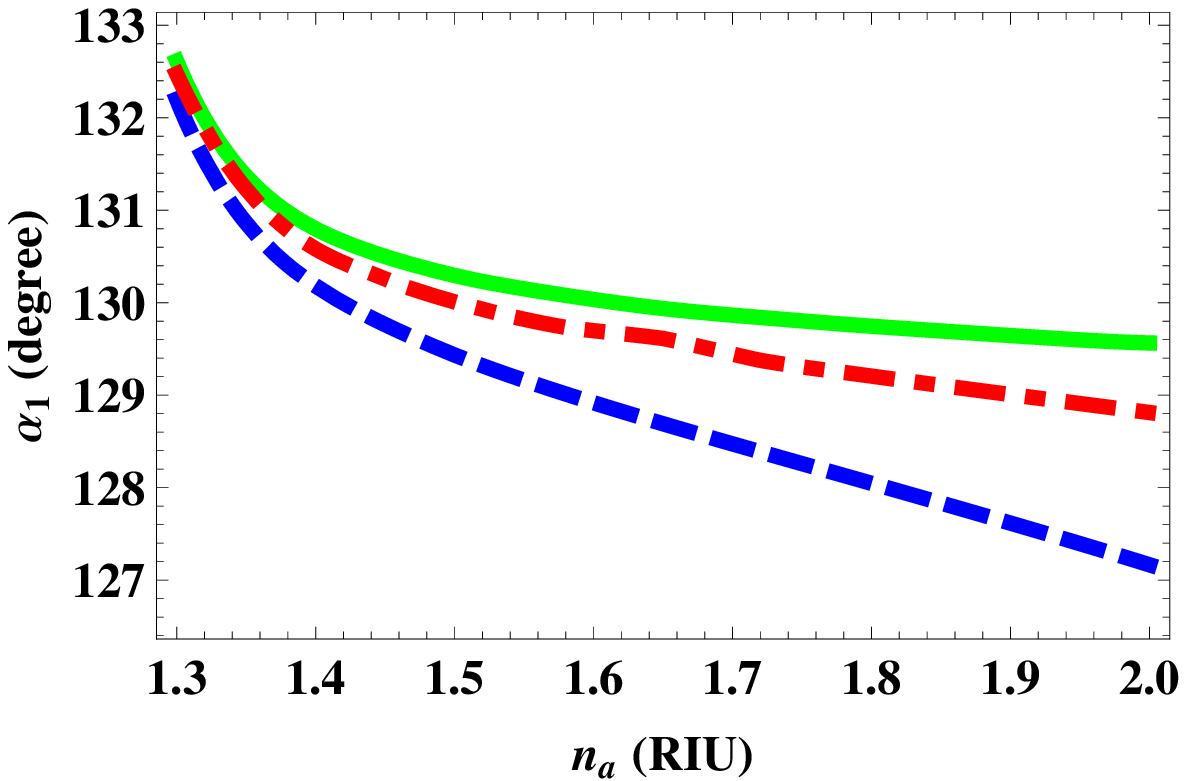,width=2.7in} \hspace{20mm}
\epsfig{file=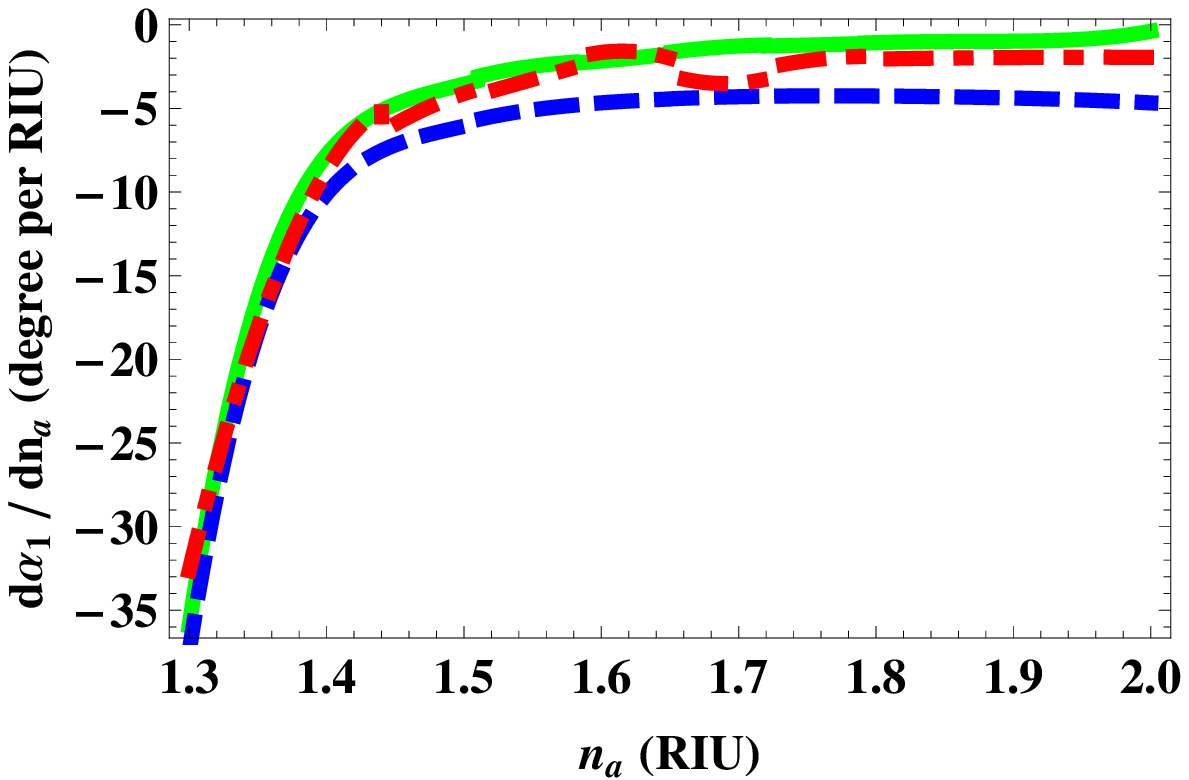,width=2.7in} \\
\epsfig{file=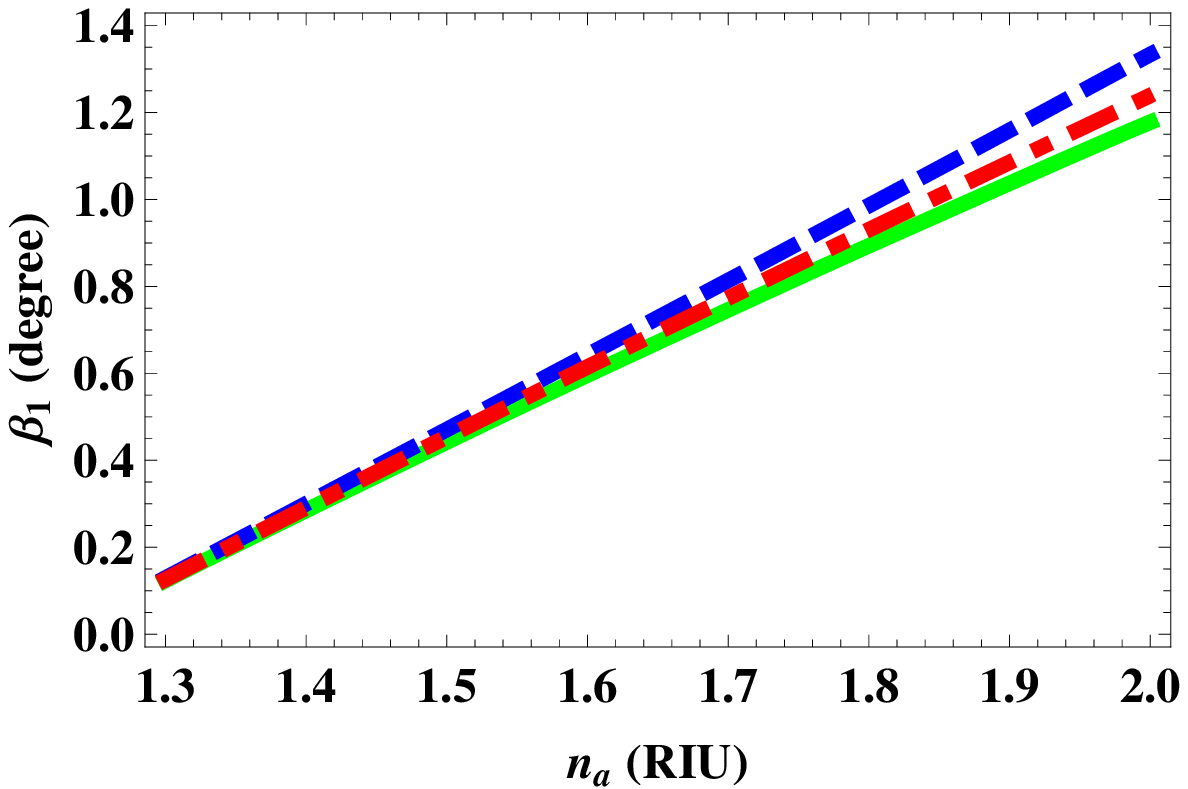,width=2.7in} \hspace{20mm}
\epsfig{file=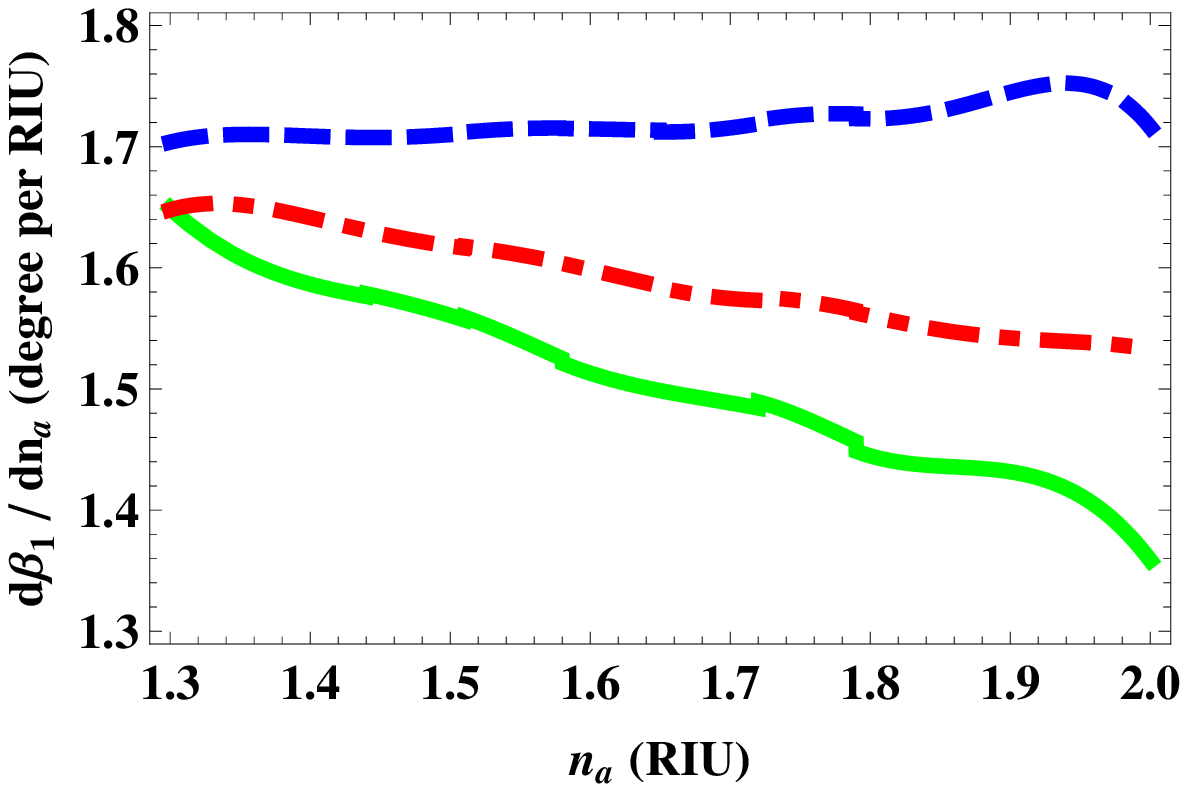,width=2.7in}
 \caption{The Euler
angles $\alpha_{1}$ and $\beta_{1}$, and the derivatives
 $d \alpha_{1} / d n_a$ and  $d \beta_{1} / d n_a$,
 plotted versus refractive index $n_a $ for
 dimensionless size  parameter values  $\ko \eta =  0.2$ (green, solid curves),
$0.3$ (red, broken dashed
curves),  and $0.4$ (blue, dashed curves).
 Values for the other component material parameters are provided in the main text.
 } \label{fig7}
\end{figure}

In the parameter regimes where the greatest sensitivities  for VWP were found , only one of the Euler angles ($\alpha$ in Figs.~\ref{fig2}--\ref{fig7}) varied sharply whereas the other Euler angle ($\beta$ in Figs.~\ref{fig2}--\ref{fig7})
remained almost constant as $n_a$ varied.
  Thus, if one were to track the directions of VWP in such parameter regimes then the tracking may only
 need to be done in one plane. This may  prove to be a helpful feature in the possible harnessing of VWP for optical sensing applications. Further studies are needed to identify practical configurations for the harnessing of VWP for optical sesning applications.

\vspace{20mm}

\section*{Appendix~1: The extended Bruggeman formalism}

The extended Bruggeman formalism is based on the nonlinear dyadic equation
\c{WLM97,M_Electromagnetics}
\begin{equation} \l{Bruggeman_eqn}
f_a \lec \le \=\eps_{\,a} -  \=\eps_{\,HCM} \ri \. \les \=I +
\=D_{\,a} \. \le \=\eps_{\,a} -  \=\eps_{\,HCM} \ri \ris^{-1} \ric +
f_b \lec \le \=\eps_{\,b} -  \=\eps_{\,HCM} \ri \. \les \=I +
\=D_{\,b} \. \le \=\eps_{\,b} -  \=\eps_{\,HCM} \ri \ris^{-1} \ric =
\=0\,.
 \end{equation}
Under the extended formalism,  the depolarization dyadics $\=D_{\,a,b}$ are viewed as
the sums \c{M_WRM}
\begin{equation}
\=D_{\,\ell} = \=D^{0}_{\,\ell} +  \=D^{+}_{\,\ell}, \qquad (\ell =
a, b),
\end{equation}
wherein the term $\=D^{0}_{\,\ell}$ represents the depolarization
contribution arising from a vanishingly small ellipsoidal particle of shape specified by
the surface dyadic  $\=U_{\,\ell}$. We have  the double integral formula
\c{M97,MW97}
\begin{equation} \l{D_o}
\=D^{0}_{\,\ell} = \frac{1}{4 \pi}\, \int^{2 \pi}_{\phi = 0}
\int^{\pi}_{\theta = 0} \frac{ \le \=U^{-1}_{\, \ell} \. \hat{\#q}
\ri \,  \le \=U^{-1}_{\, \ell} \. \hat{\#q} \ri \; \sin \theta}{\le
\=U^{-1}_{\, \ell} \. \hat{\#q} \ri \. \=\eps_{\, HCM} \. \le
\=U^{-1}_{\, \ell} \. \hat{\#q} \ri}\, d \theta\, d \phi \, , \qquad
(\ell = a, b),
\end{equation}
with the unit vector $\hat{\#q} = \sin \theta \cos \phi \,
\hat{\#x} + \sin \theta \sin \phi \, \hat{\#y} + \cos \theta \,
\hat{\#z}$. The contribution to the depolarization dyadic arising specifically from the
nonzero size of the component particles is represented by the
 dyadic term $\=D^{+}_{\,\ell}$. It is convenient to express this dyadic
 as a subdyadic of the  6$\times$6 dyadic $\underline{\underline{\mathbb{D}}}^{+}_{\,\ell}$, as defined via
\begin{equation}
\les \, \=D^{+}_{\,\ell} \, \ris_{m n} = \les \,
\underline{\underline{\mathbb{D}}}^{+}_{\,\ell} \, \ris_{m n},
\qquad \le m, n \in \lec 1, 2, 3 \ric \ri.
\end{equation}
Here \c{M_WRM}
\begin{eqnarray} \l{D_plus}
\underline{\underline{\mathbb{D}}}^{+}_{\,\ell} &=&
\frac{\omega^4}{4 \pi \muo } \int^{2 \pi}_{\phi = 0}
\int^{\pi}_{\theta = 0} \frac{\sin \theta}{\les \le \=U^{-1}_{\,
\ell} \. \hat{\#q} \ri \. \=\eps_{\, HCM} \. \le \=U^{-1}_{\, \ell}
\. \hat{\#q} \ri \ris \le \=U^{-1}_{\, \ell} \. \hat{\#q} \ri \. \le
\=U^{-1}_{\, \ell} \. \hat{\#q} \ri } \nonumber \\ && \times \Bigg[
\frac{1}{ \kappa_+ - \kappa_-  }  \Bigg( \frac{\exp \le i \eta q \ri
}{2 q^2} \le 1 - i \eta q\ri
 \Big\{ \,
 \mbox{det} \les \underline{\underline{\mathbb{A}}} (\=U^{-1}_{\, \ell} \.\#q ) \ris \, \underline{\underline{\mathbb{G}}}^{+} (\=U^{-1}_{\,
\ell} \.\#q) \nonumber \\
&&
  +  \mbox{det} \les \underline{\underline{\mathbb{A}}} (-\=U^{-1}_{\, \ell} \.\#q ) \ris  \,  \underline{\underline{\mathbb{G}}}^{+} (-\=U^{-1}\.\#q
) \Big\} \Bigg)^{q=\sqrt{ \kappa_+ }}_{q=\sqrt{ \kappa_- }} + \frac{
 \mbox{det} \les \underline{\underline{\mathbb{A}}} (\#0 ) \ris} {\kappa_+  \, \kappa_- }\,
\underline{\underline{\mathbb{G}}}^{+} (\#0)
 \Bigg] \; d \theta \; d \phi, \qquad (\ell = a,b), \nonumber \\ &&
\end{eqnarray}
wherein  $\kappa_\pm $ are the $q^2$ roots of $\mbox{det} \les
\underline{\underline{\mathbb{A}}}(\=U^{-1}\.\#q) \ris = 0$, the
vector $\#q = q \, \hat{\#q}$, and the 6$\times$6 dyadics
\begin{equation}
\underline{\underline{\mathbb{A}}} (\#p)  = \les \begin{array}{cc}
\=\eps_{\,HCM} & \le \#p / \omega \ri \times \=I \vspace{2mm}  \\
-\le \#p / \omega \ri \times \=I  & \muo \, \=I
\end{array}
 \ris
\end{equation}
and
\begin{equation}
\underline{\underline{\mathbb{G}}}^{+} (\#p) =
\underline{\underline{\mathbb{A}}}^{-1} (\#p) - \lim_{| \#p | \to
\infty} \underline{\underline{\mathbb{A}}}^{-1} (\#p) .
\end{equation}

The HCM permittivity dyadic
$\=\eps_{\,HCM}$ may be extracted from the nonlinear dyadic equation~\r{Bruggeman_eqn} by standard numerical techniques, such as the Jacobi technique
\c{Jacobi}.

\vspace{20mm}

\section*{Appendix~2: Estimates of the HCM's constitutive parameters}

\begin{figure}[!h]
\centering \psfull \epsfig{file=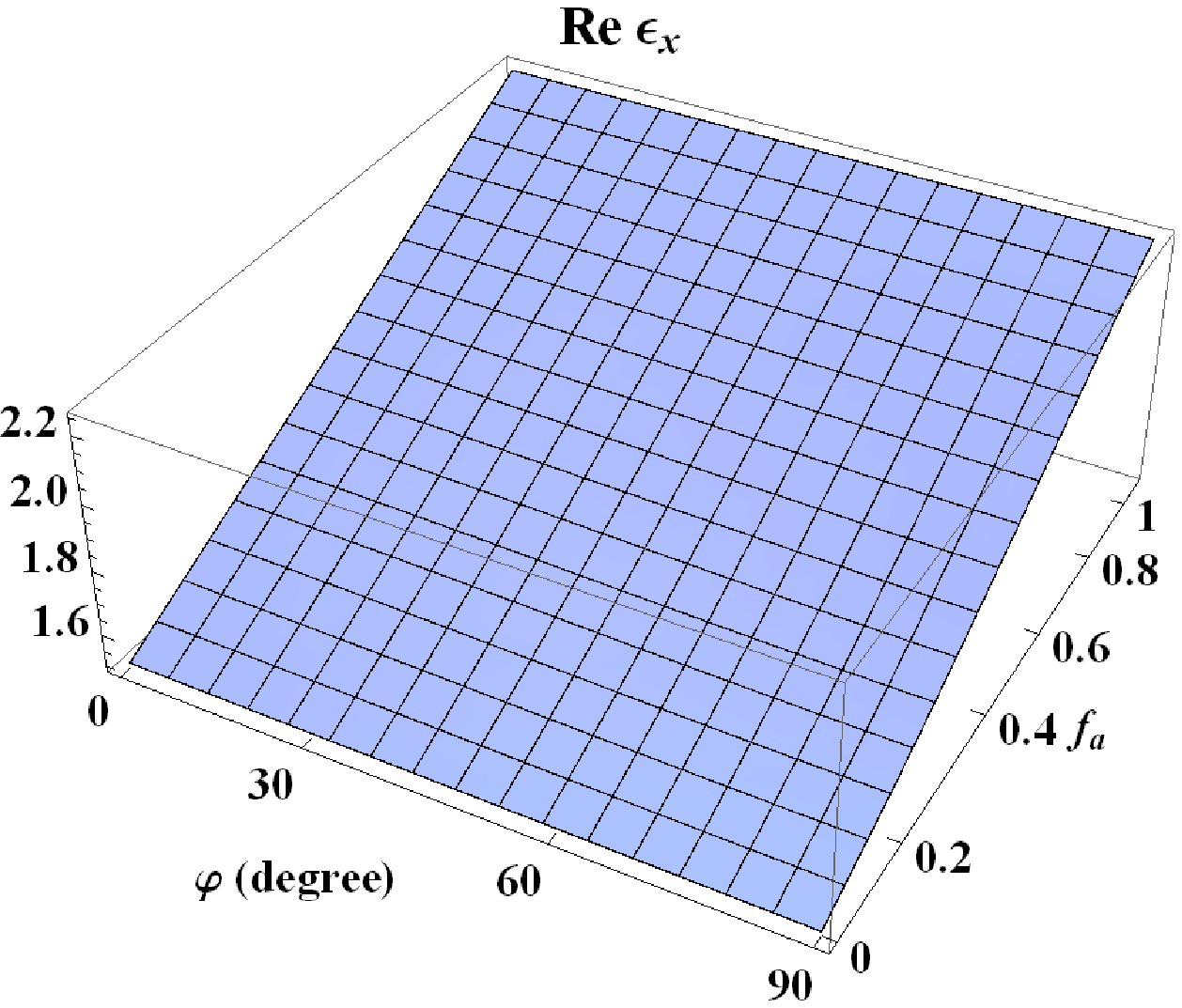,width=2.2in}
\hspace{20mm}
\epsfig{file=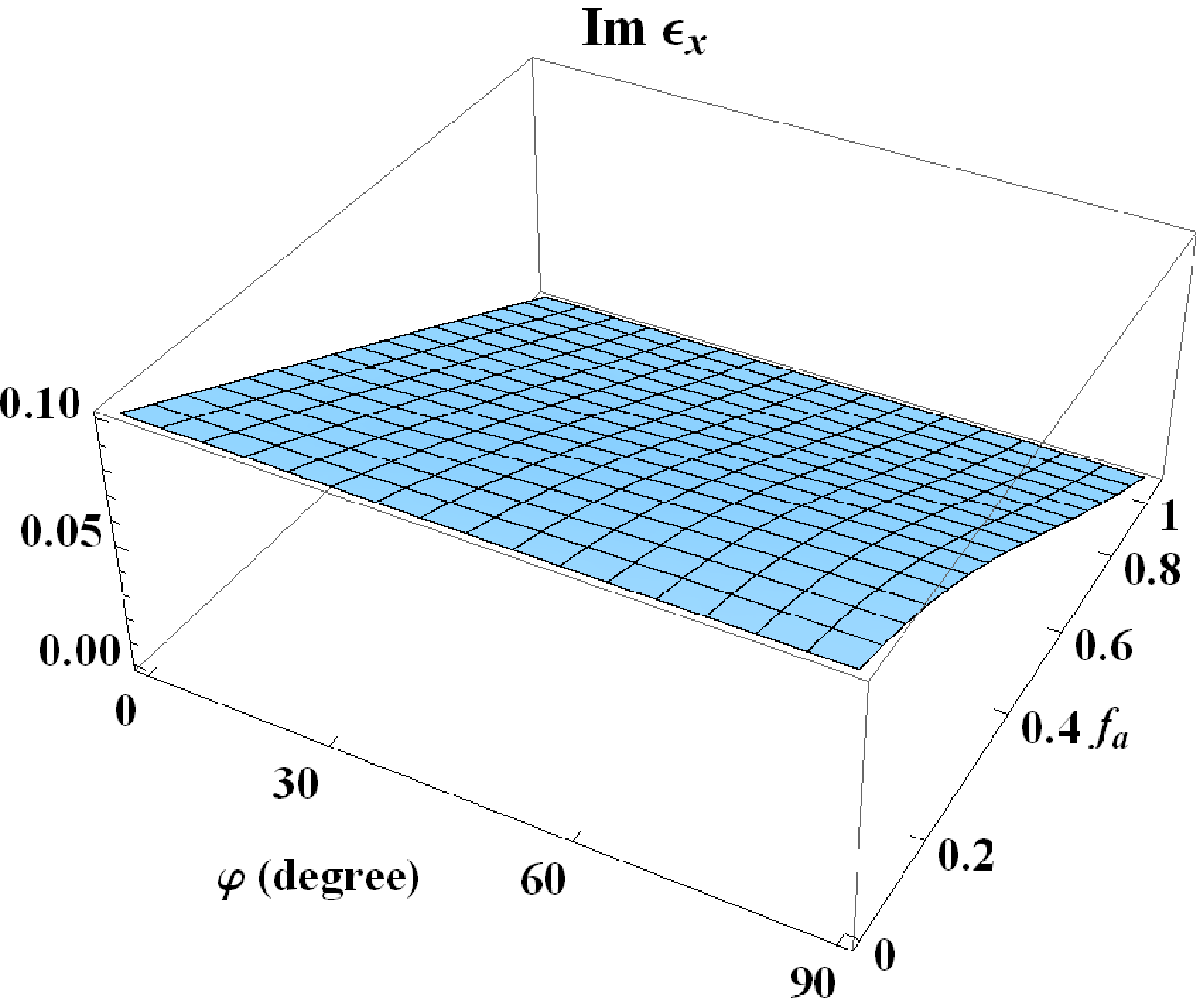,width=2.2in} \\
\epsfig{file=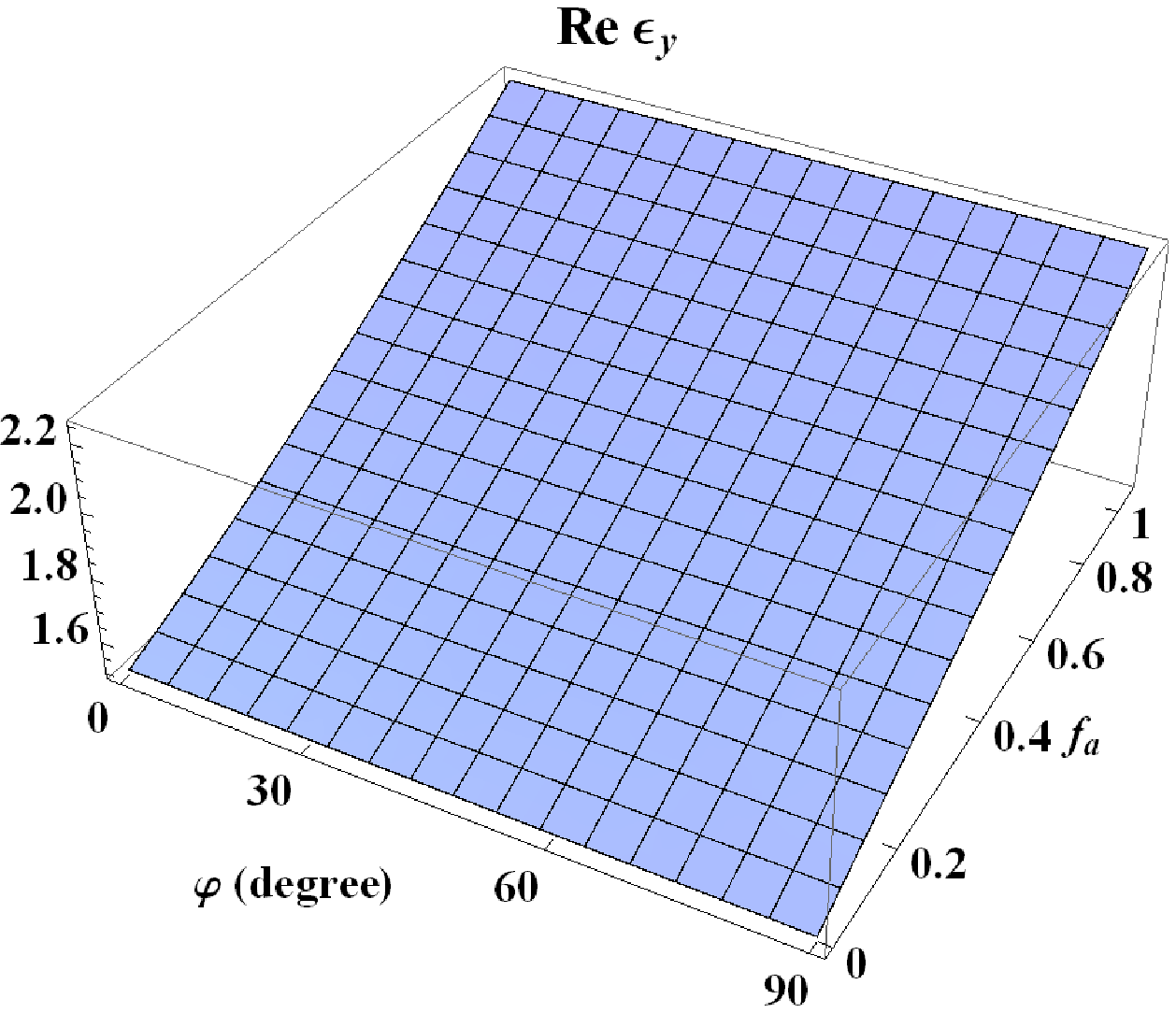,width=2.2in} \hspace{20mm}
\epsfig{file=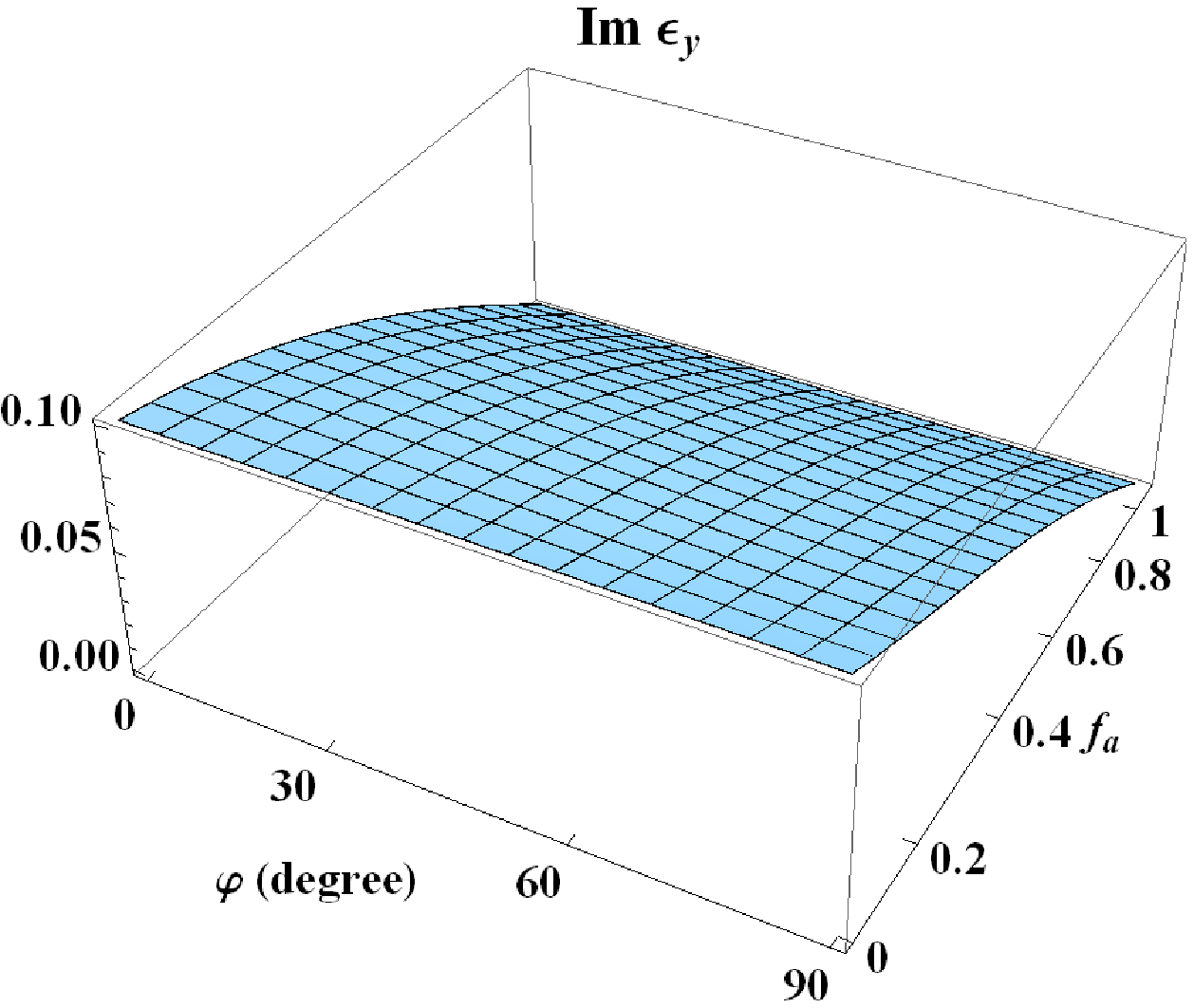,width=2.2in} \\
\epsfig{file=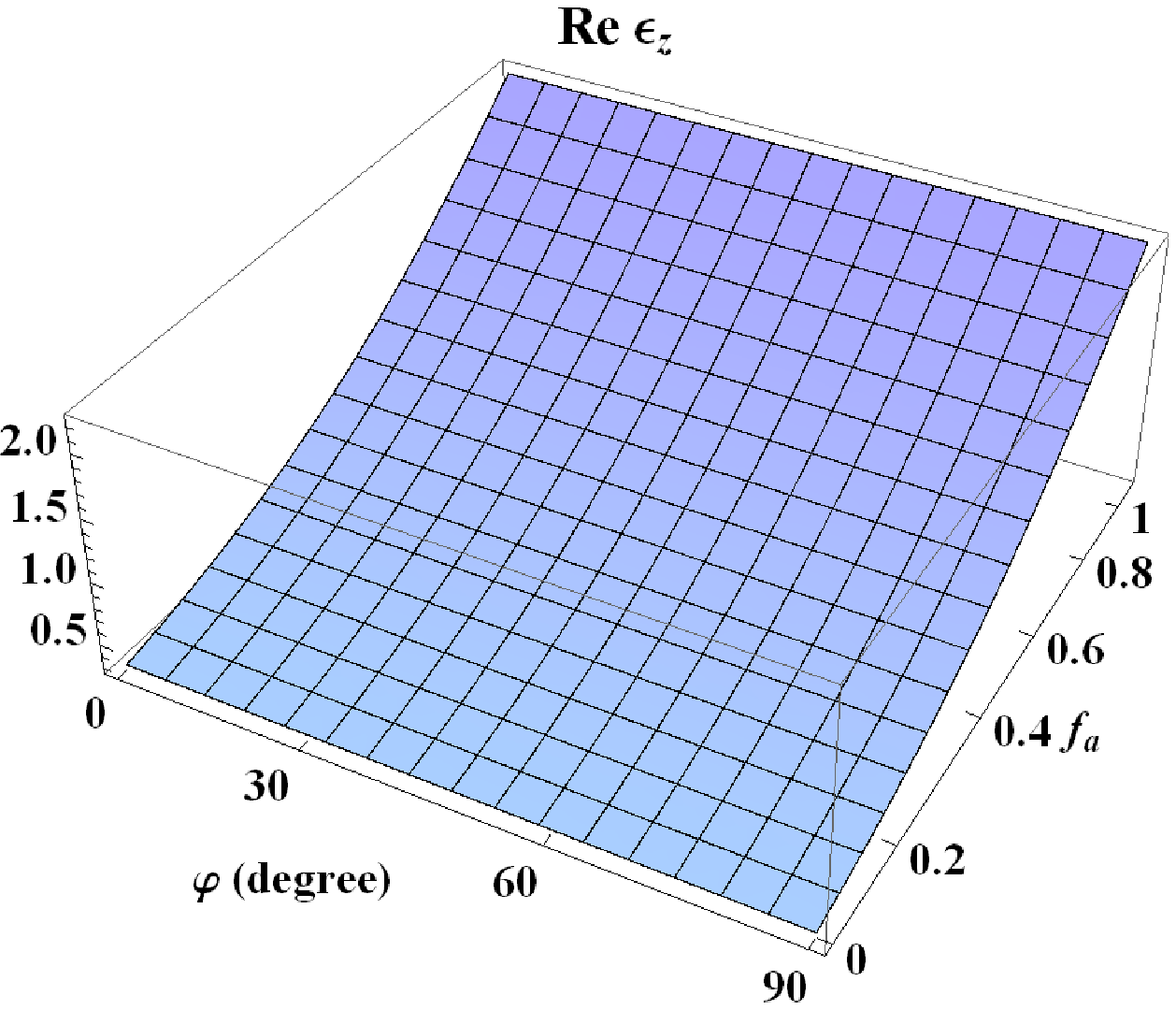,width=2.2in} \hspace{20mm}
\epsfig{file=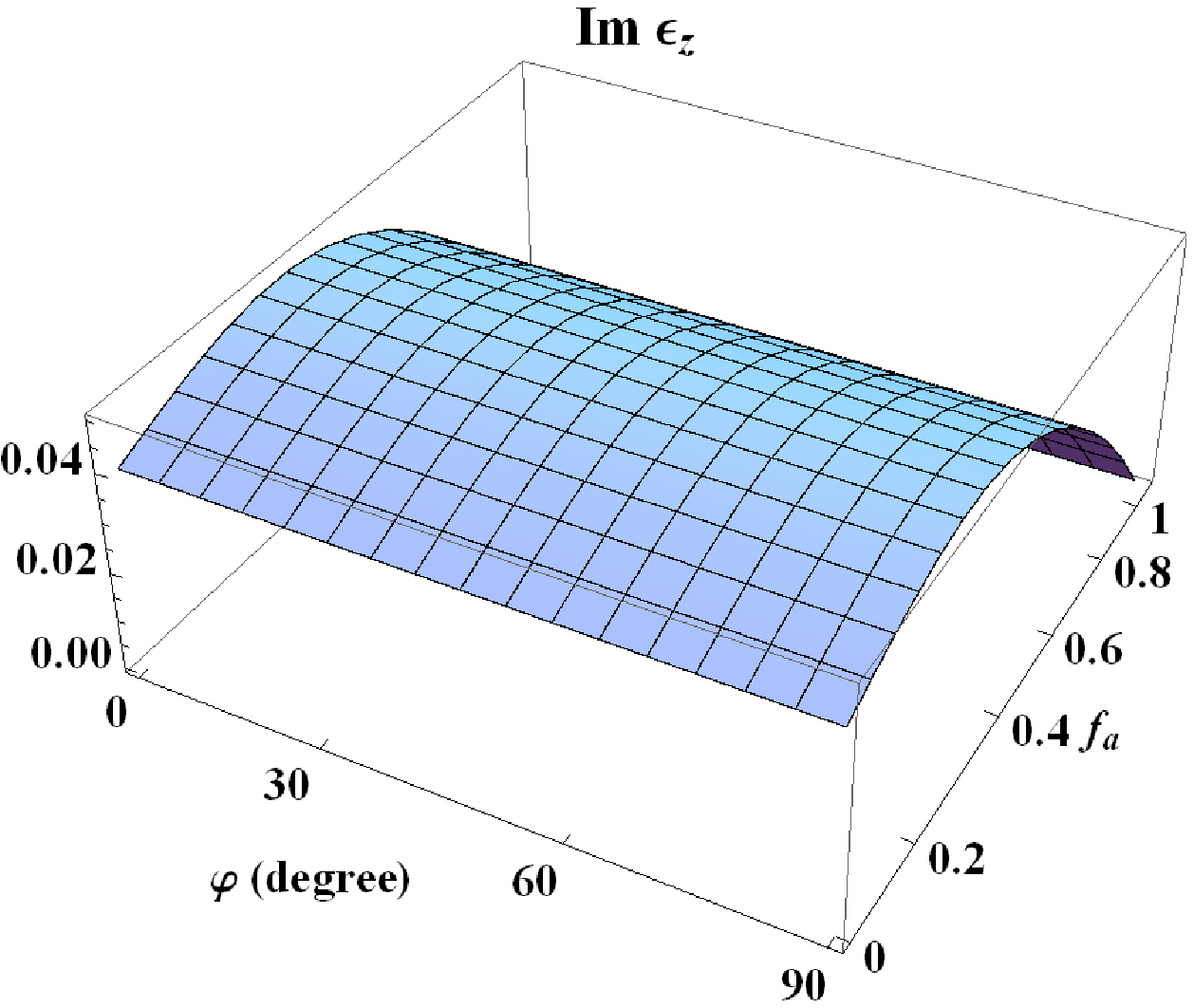,width=2.2in} \\
\epsfig{file=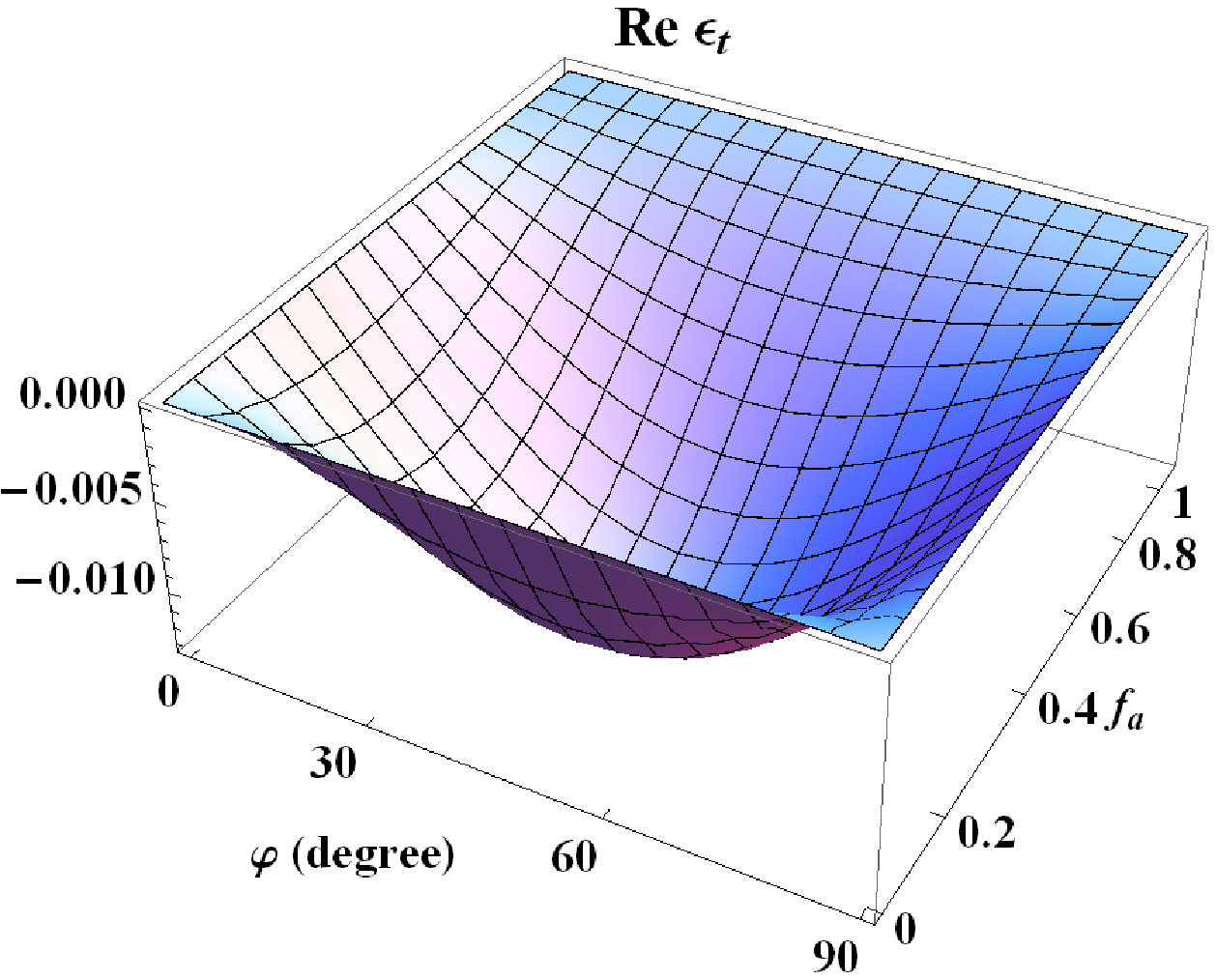,width=2.2in} \hspace{20mm}
\epsfig{file=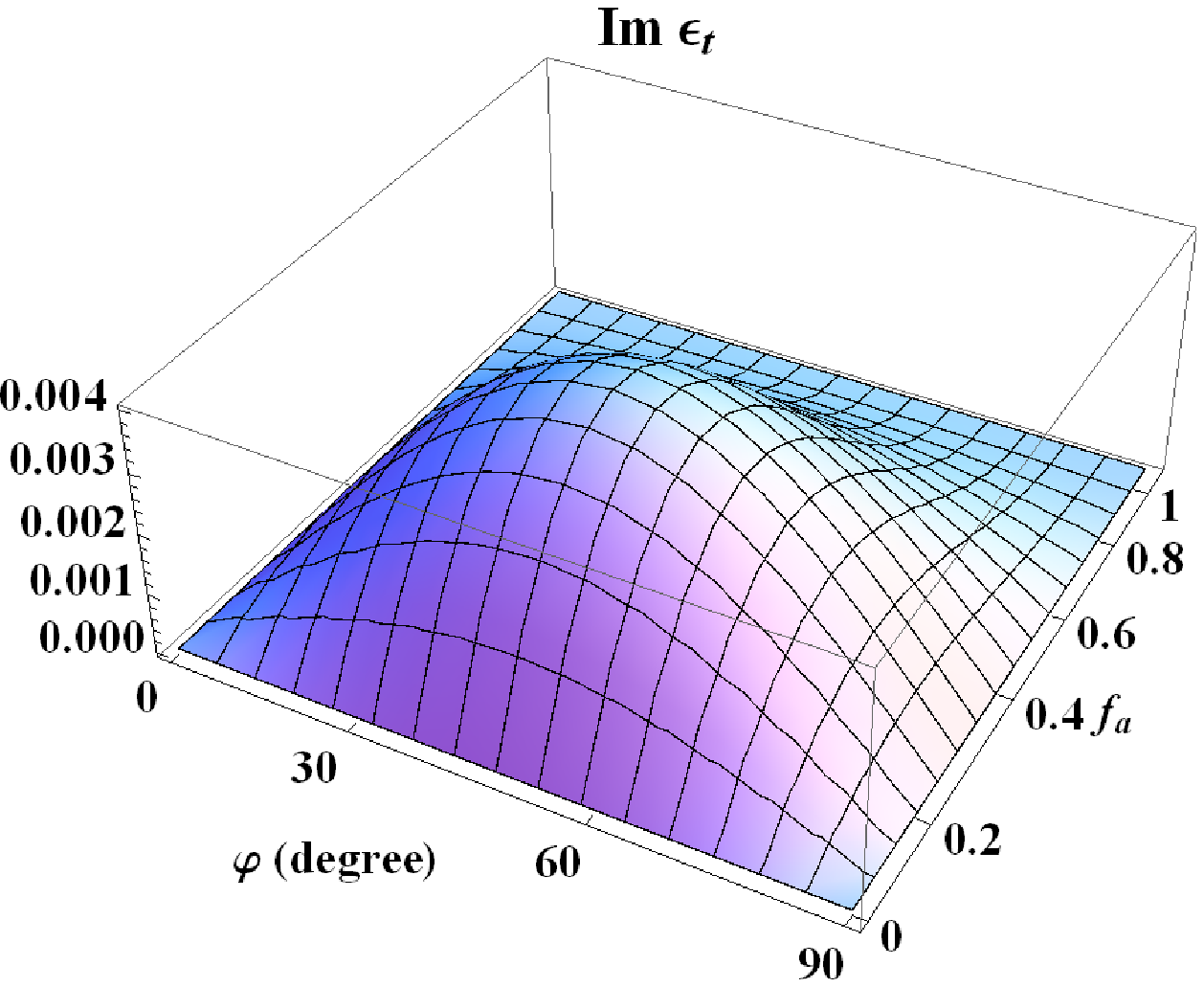,width=2.2in}
 \caption{Real and imaginary parts of the extended Bruggeman estimates of relative permittivity parameters of the HCM plotted versus volume fraction $f_a \in \le 0, 1 \ri$
 and particle orientation angle $\varphi \in \le 0^\circ, 90^\circ \ri$.
The dimensionless size parameter $\ko \eta = 0.2$, the dissipation parameter $\delta = 0.1$, the anisotropy parameter $\tau = 1$,  and the particle eccentricity
parameter $\rho = 9$. } \label{figA1}
\end{figure}

\begin{figure}[!h]
\centering \psfull \epsfig{file=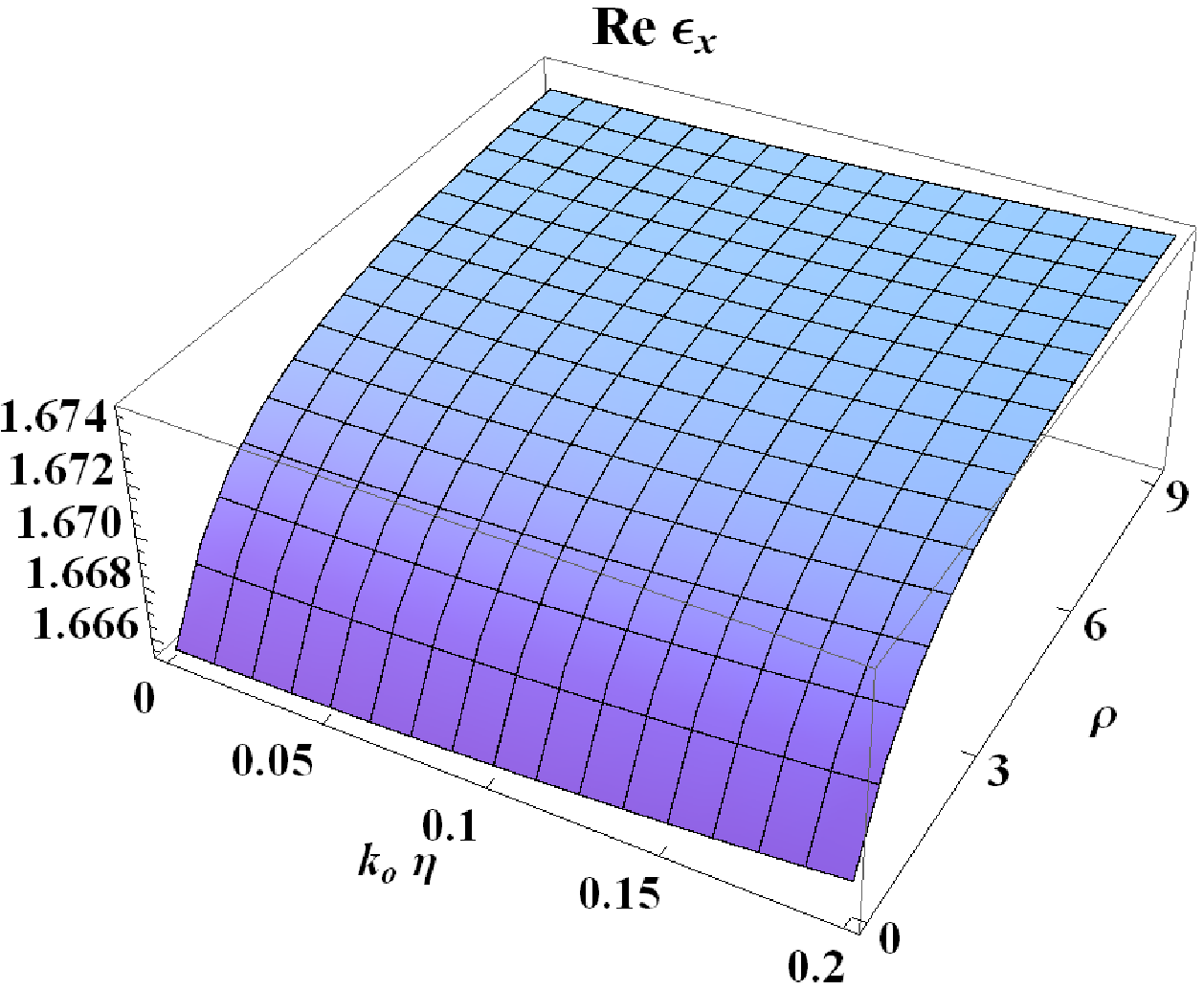,width=2.2in}
\hspace{20mm}
\epsfig{file=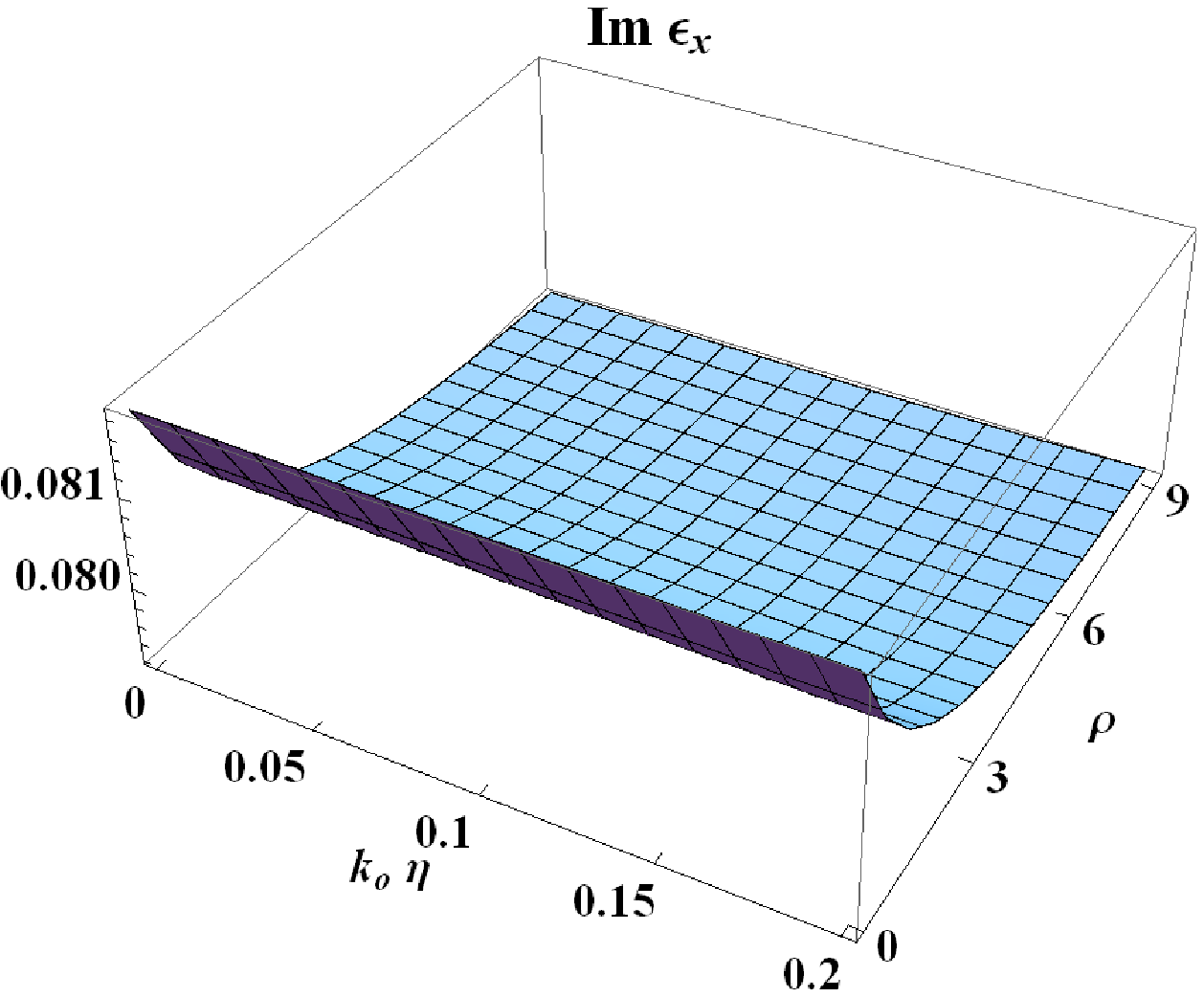,width=2.2in} \\
\epsfig{file=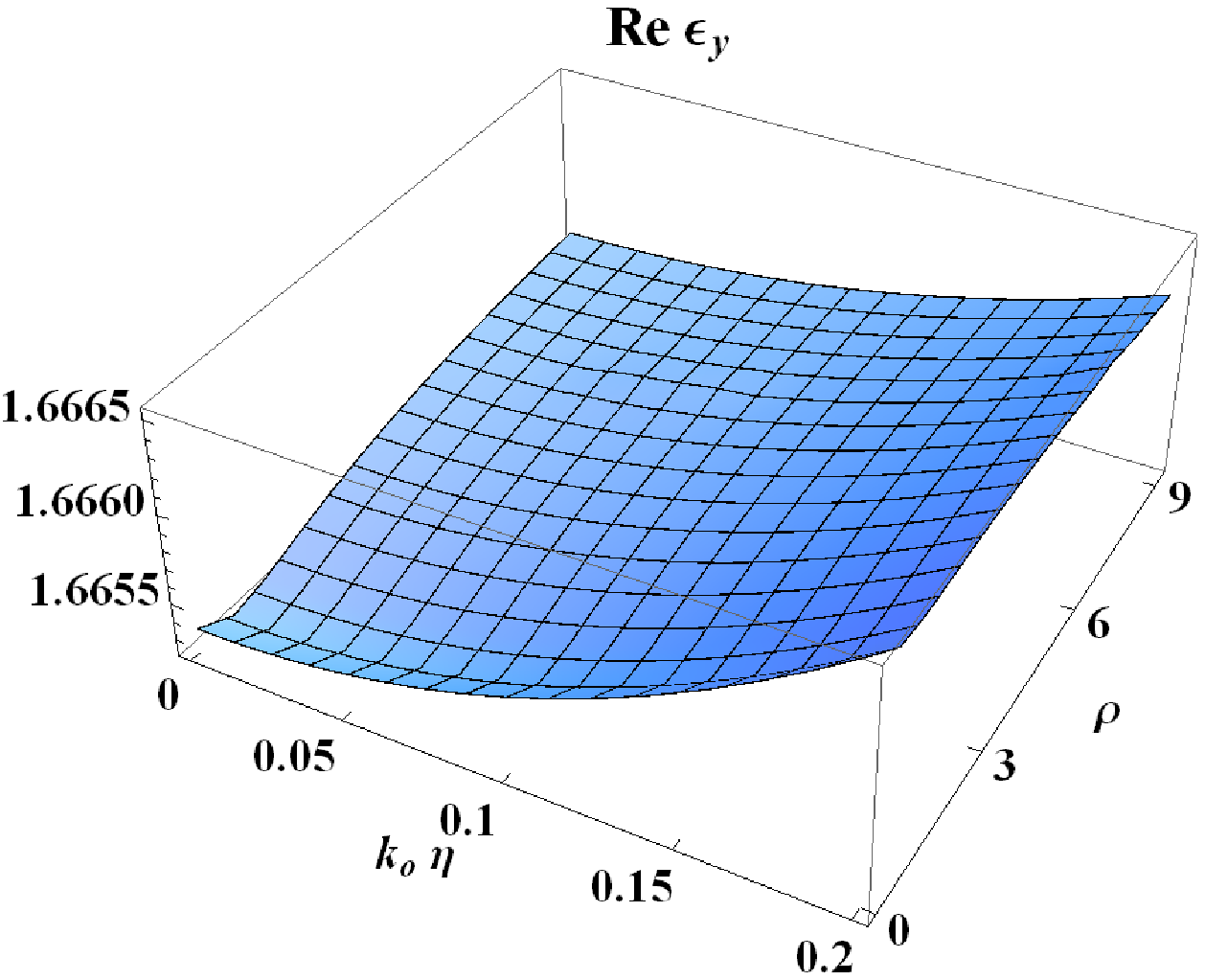,width=2.2in} \hspace{20mm}
\epsfig{file=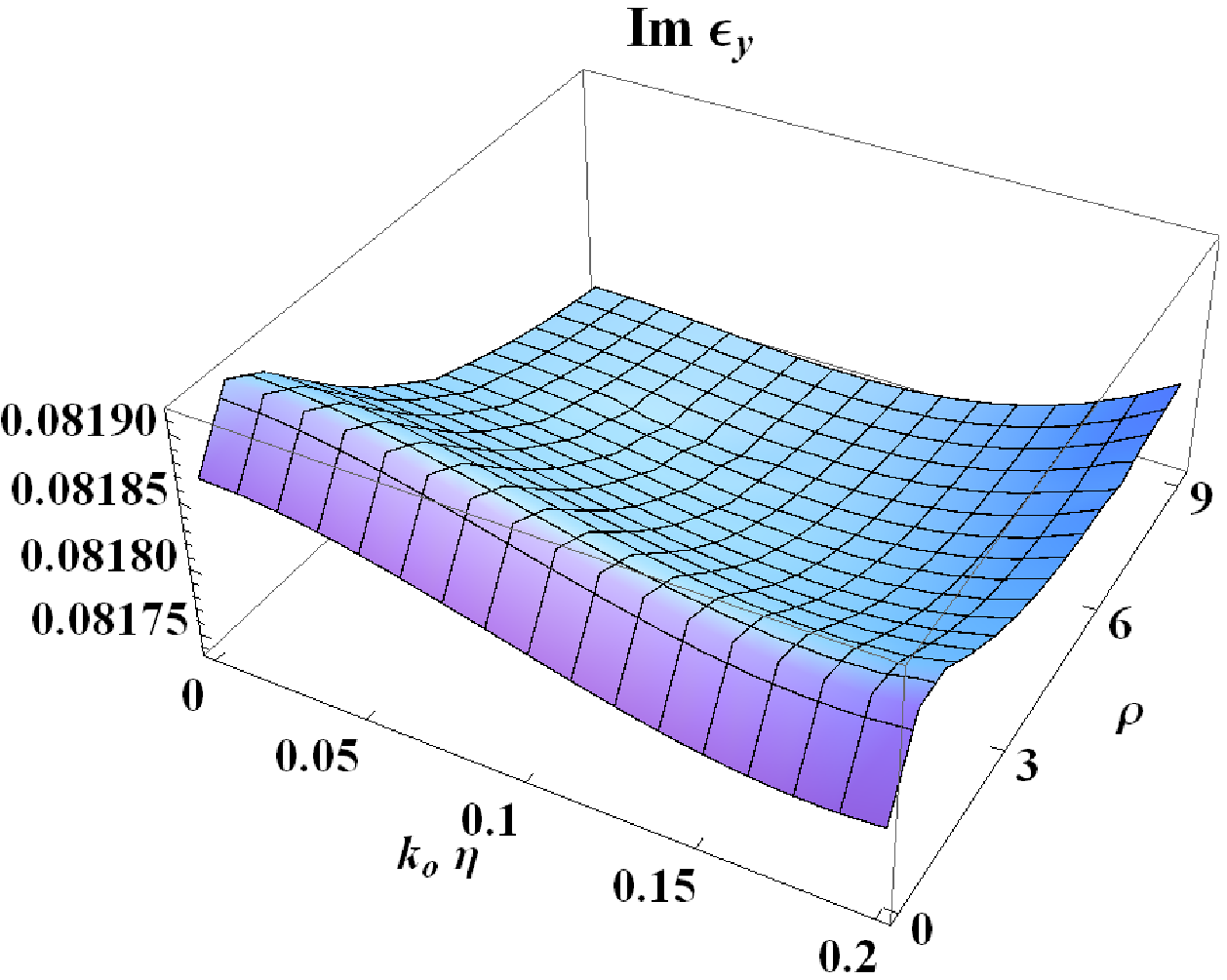,width=2.2in} \\
\epsfig{file=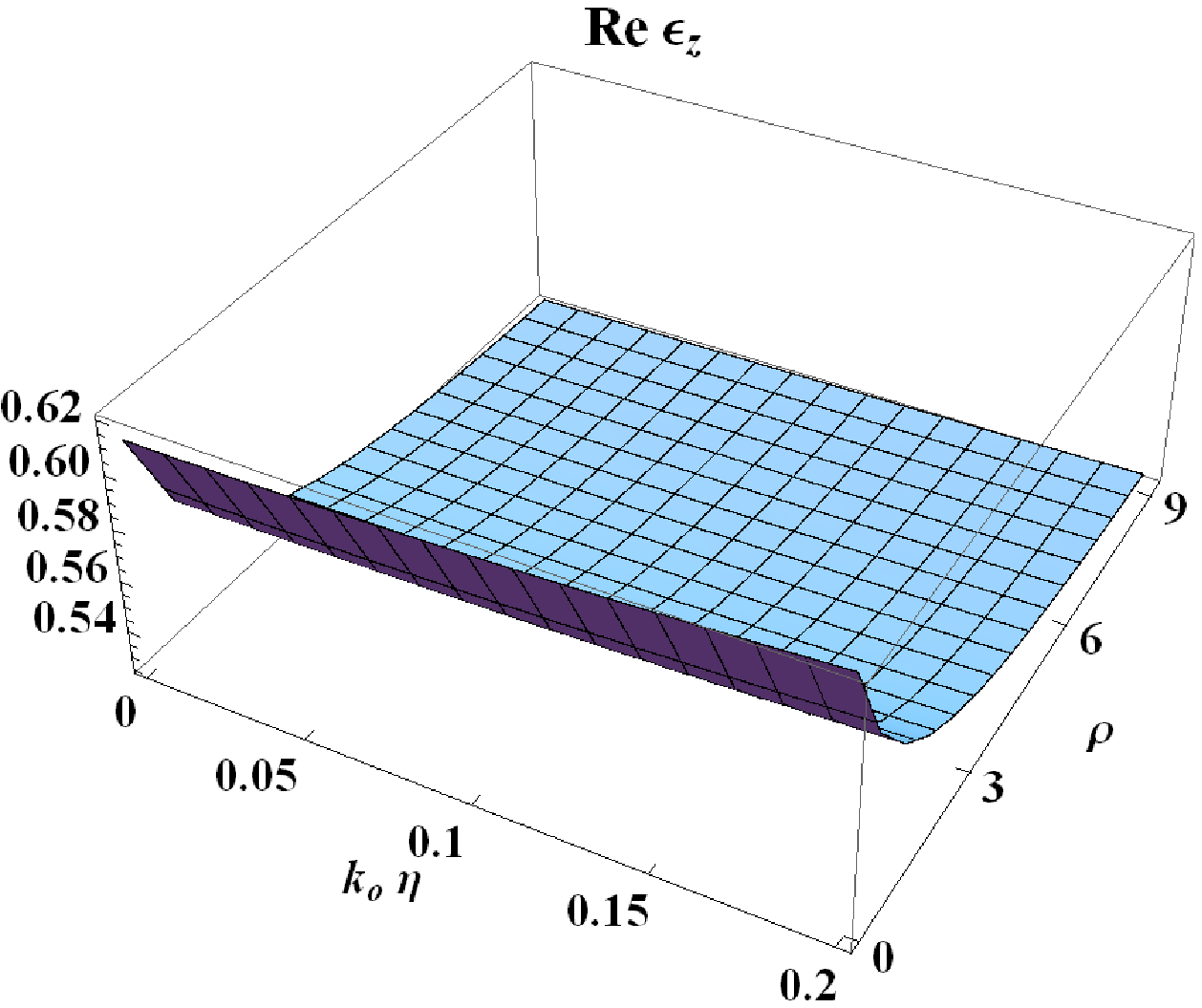,width=2.2in} \hspace{20mm}
\epsfig{file=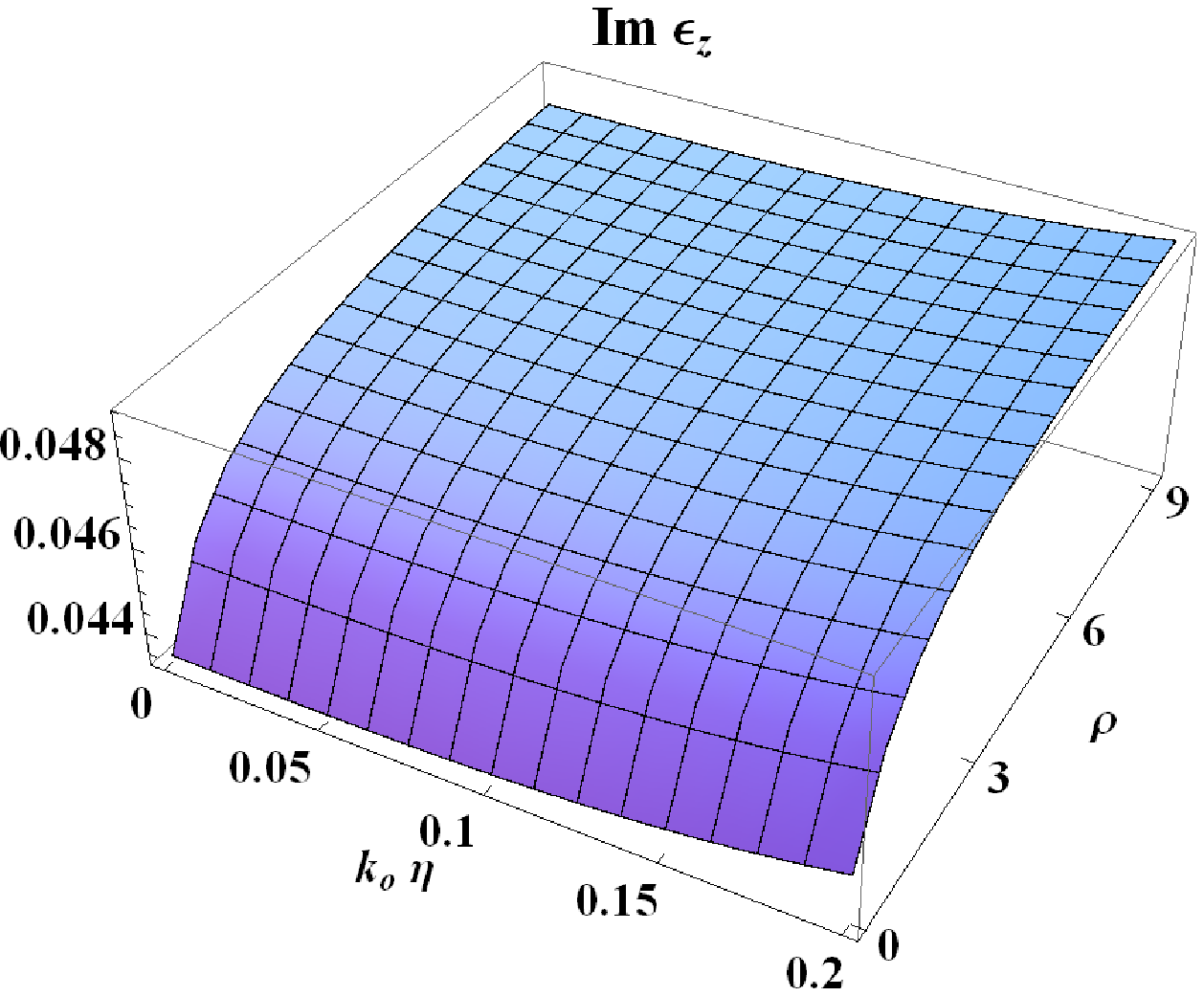,width=2.2in} \\
\epsfig{file=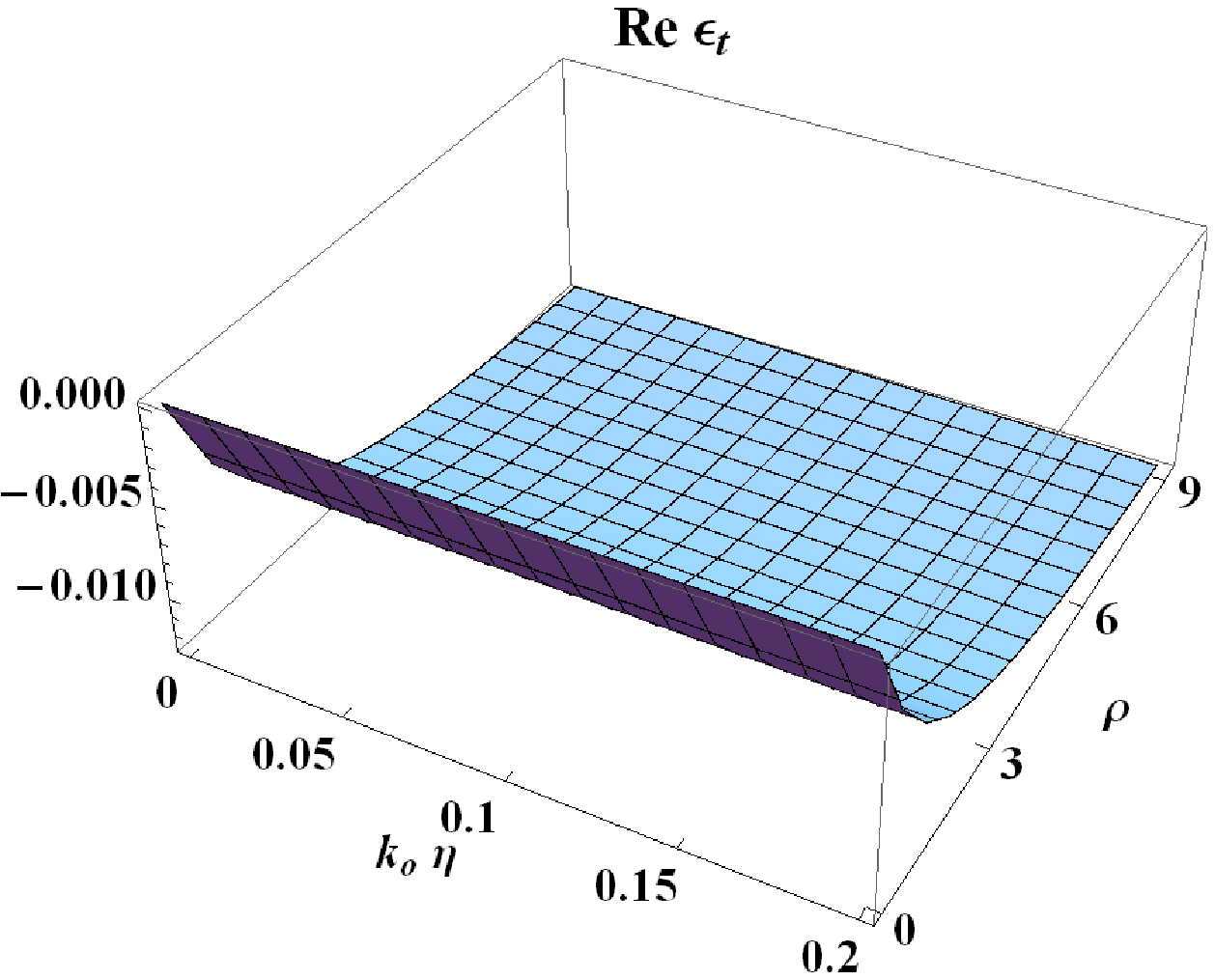,width=2.2in} \hspace{20mm}
\epsfig{file=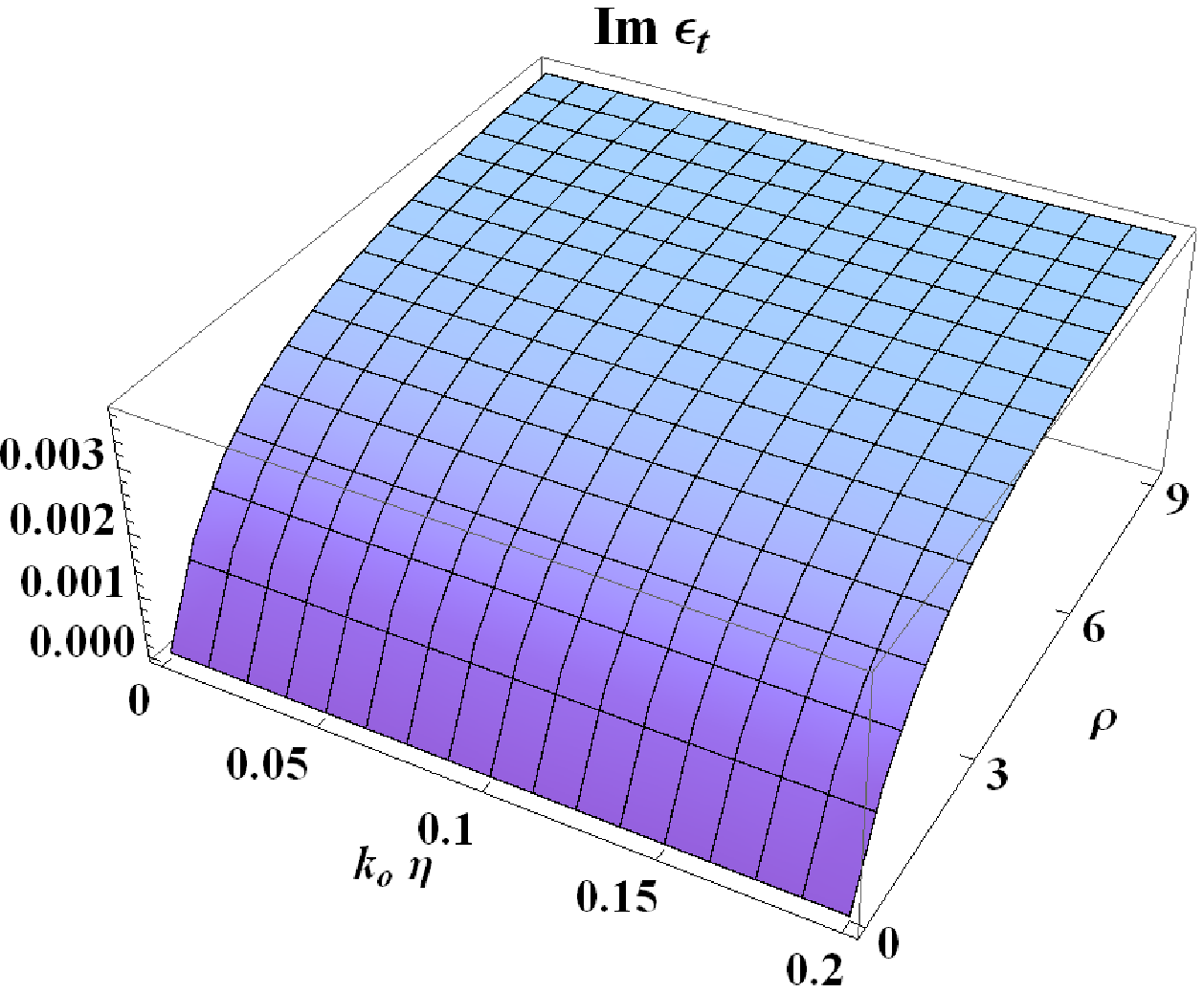,width=2.2in}
 \caption{As Fig.~\ref{figA1} except that the HCM's relative permittivity
 parameters are plotted versus the dimensionless size parameter $\ko \eta \in \le 0,  0.2 \ri$ and the particle eccentricity
parameter $\rho \in \le 0, 9 \ri$. The volume fraction $f_a = 0.25 $
 and particle orientation angle $\varphi = 45^\circ $.} \label{figA2}
\end{figure}

Here estimates of the extended Bruggeman estimates of the HCM's relative permittivity
parameters $\eps_{x,y,z,t}$ are presented. Component material $a$  is specified by $n_a = 1.5$
while component material $b$ is specified by the anisotropy parameter $\tau = 1$ and dissipation parameter $\delta = 0.1$, per Eqns.~\r{components}.
In
Fig.~\ref{figA1}, the real and imaginary parts of $\eps_{x,y,z,t}$ are plotted as functions of particle
orientation angle $\varphi$ and volume fraction $f_a$; for these calculations the dimensionless size parameter $\ko \eta
= 0.2$ and the particle eccentricity parameter $\rho = 9$.
In Fig.~\ref{figA2}, the real and imaginary parts of $\eps_{x,y,z,t}$ are plotted against
dimensionless  size parameter $\ko \eta$ and the particle eccentricity parameter
$\rho$; for these calculations  the volume fraction $ f_a = 0.25$ and the particle
 orientation angle $\varphi = 45^\circ$. For comprehensive discussions on the relationships between the
 extended (and non-extended)  Bruggeman estimates  of the HCM's relative permittivity
parameters and the parameters which characterize the component materials, the reader is referred to
 earlier works
\c{ML03,ML_WRM,M2011_JOPA}.

\newpage

\newpage

\newpage

\newpage

\newpage

\newpage

\newpage

\newpage

\newpage

\end{document}